\documentclass[aps,prl,twocolumn,showpacs,amsmath,amssymb,floatfix]{revtex4-1}

\usepackage{graphicx,color,framed}
\usepackage{hyperref}
\usepackage{times}
\usepackage{enumerate}
\usepackage{lipsum}
\usepackage{slashed}
\usepackage{array}
\usepackage{textcomp}
\usepackage{amsthm}

\hypersetup{
    colorlinks=true, 
    linktoc=all,     
    linkcolor=blue,  
}

\newtheorem{thm}{Theorem}

\newtheorem{lemma}{Lemma}
\newtheorem*{lemma*}{Lemma}
\newtheorem{claim}{Claim}

\def \beq {\begin{equation}}
\def \eeq {\end{equation}}
\def \beqa {\begin{eqnarray}}
\def \eeqa {\end{eqnarray}}
\def \bseq {\begin{subequations}}
\def \eseq {\end{subequations}}
\newcommand \dg {\dagger}

\newcommand \ran {\rangle}
\newcommand \lan {\langle}

\newcommand \ep {\epsilon}

\newcommand \nnb {\nonumber}
\newcommand \vphi {\varphi}

\newcommand \ch {\hat{c}}
\newcommand \nh {\hat{n}}
\newcommand \ph {\hat{\phi}}

\begin{document}

\title{Rigorous results on topological superconductivity with particle number conservation}

\author{Matthew F. Lapa}
\email[email address: ]{mlapa@uchicago.edu}
\affiliation{Kadanoff Center for Theoretical Physics, University of Chicago, Chicago, IL, 60637, USA}

\author{Michael Levin}
\email[email address: ]{malevin@uchicago.edu}
\affiliation{Kadanoff Center for Theoretical Physics, University of Chicago, Chicago, IL, 60637, USA}


\begin{abstract}

Most theoretical studies of topological superconductors and Majorana-based quantum 
computation rely on a mean-field approach to describe superconductivity. A potential problem with this approach is that real superconductors are described by \emph{number-conserving} Hamiltonians 
with \emph{long-range} interactions, so their topological properties may not be correctly captured by mean-field models that violate number conservation and have short-range interactions.
To resolve this issue, reliable results on number-conserving models of superconductivity are essential. As a first step in this direction, we use rigorous methods to study a number-conserving toy model of a topological superconducting wire. 
We prove that this model exhibits many of the desired properties of the mean-field models, including a 
finite energy gap in a sector of fixed total particle number, the existence of long range “Majorana-like”
correlations between the ends of an open wire, and a change in the ground state
fermion parity for periodic vs. anti-periodic boundary conditions. These results show that many of the remarkable properties of mean-field models of topological superconductivity persist in more realistic models with number-conserving dynamics.

\end{abstract}


\maketitle

\textit{Introduction:} Topological superconductors (TSCs) hosting unpaired Majorana 
fermions~\cite{kitaev,read-green} are of great 
interest as a potential platform for fault-tolerant quantum computation. However, most 
studies of TSCs rely on a mean-field (or Bogoliubov-de Gennes) approach to
describe superconductivity. Recently, A.J. 
Leggett and others have questioned the validity of this 
approach for quantum computing applications~\cite{Leggett1,Leggett2,ortiz2,lin-leggett1,lin-leggett2}.
A key concern is that real superconductors are described by 
\emph{number-conserving} Hamiltonians with
\emph{long-range} interactions, so their topological properties may not be correctly captured by mean-field Hamiltonians that violate number-conservation and have short-range interactions.

To resolve this issue, we need reliable results on number-conserving models of topological superconductors.
In this paper we provide such results for a toy model of a one-dimensional (1D)
TSC. We rigorously prove that this model possesses many of the desired 
properties of the mean-field models, including a finite energy gap in a sector of fixed total particle 
number, long range ``Majorana-like'' correlations between the ends of an open wire, and
a change in the ground state fermion parity for 
periodic vs. anti-periodic boundary conditions (a ``fermion parity switch''). 
In addition, our model captures some qualitative features of \emph{proximity-induced}
superconductivity, and so our results are relevant for current experimental platforms 
for TSCs~\cite{PhysRevLett.105.077001,PhysRevLett.105.177002,PhysRevB.88.020407,
mourik2012signatures,das2012zero,rokhinson2012fractional,
deng2012anomalous,PhysRevLett.110.126406,PhysRevB.87.241401,yazdani2014observation,albrecht2016exponential}.

The problem of number conservation and topological superconductivity has been investigated previously 
using several approaches including bosonization~\cite{cheng-tu,
fidkowski2011,Sau2011,tsvelik2011zero,cheng-lutchyn,KSH,knapp2019number,RuhmanBergAltman} and exactly solvable 
models~\cite{PhysRevB.82.224510,ortiz1,ortiz2,iemini2015localized,lang2015topological,
PhysRevB.96.115110} (for additional approaches see 
\cite{Kraus2013,burnell2018,RA,lin-leggett1,lin-leggett2,cui2019}).
The first approach gives general results on low-energy field theory models, but
requires various approximations. The second approach gives rigorous results on concrete 
microscopic models, but only for a fine-tuned set of parameter values. 
Our approach differs from these previous works in several ways. Most importantly, we provide 
rigorous results on a concrete microscopic model, \emph{and} our results hold for a wide range of 
parameter values (no fine-tuning). 
In addition, our model has true superconducting order, as opposed to the 1D 
quasi-long range order present in 
Refs.~\onlinecite{cheng-tu,fidkowski2011,Sau2011,tsvelik2011zero,cheng-lutchyn,KSH,knapp2019number,cui2019}.

\textit{Review of the mean-field model:} We start with a review of the standard mean-field model of a 
1D TSC: the Kitaev p-wave wire model~\cite{kitaev}.
This model consists of spinless fermions $\ch_j$ on the sites $j\in\{1,\dots,L\}$ of a wire of length $L$,
and the Hamiltonian with open boundary conditions takes the form
\begin{align}
	\hat{H}_{\text{MF}} =& -\frac{t}{2}\sum_{j=1}^{L-1}(\ch^{\dg}_j\ch_{j+1}+\text{h.c.}) - \frac{\Delta}{2}\sum_{j=1}^{L-1}(\ch_j\ch_{j+1}+\text{h.c.}) \nnb \\
	-&\ \mu  \left(\hat{N}_w -\frac{L}{2}\right)\ ,
\end{align}
where $\hat{N}_w=\sum_{j=1}^L \ch^{\dg}_j \ch_j$ is the total number operator (``w'' stands for ``wire''), 
$t>0$ and $\Delta$ are energies for nearest neighbor hopping and pairing, respectively,
and $\mu$ is the chemical potential. This model is in its topological phase for 
$|\mu| < t$ and $\Delta \neq 0$. In this phase, and with open boundary conditions, 
$\hat{H}_{\text{MF}}$ has a unique ground state and a finite energy gap in each sector of fixed 
fermion parity. The two lowest energy states with opposite fermion parity are nearly degenerate 
(their splitting is exponentially small in $L$~\cite{kitaev,bravyi-koenig}), and this degeneracy is related 
to the presence of Majorana fermions localized near the ends of the wire. 

The properties of this model are easiest to see at the special point $t=\Delta$ and $\mu=0$. 
At this point $\hat{H}_{\text{MF}}$ commutes with the Majorana fermion operators 
$\hat{a}_1 = -i (\ch_1 -\ch^{\dg}_1)$ and $\hat{b}_L = \ch_L+\ch^{\dg}_L$ localized on sites $1$ and $L$, 
respectively. As a consequence, $\hat{H}_{\text{MF}}$ has two degenerate ground states $|\pm\ran$ 
(with fermion parity $\pm 1$) that satisfy $i\hat{a}_1\hat{b}_L |\pm\ran = \pm|\pm\ran$.
This property implies the existence of a \emph{long range correlation}
\beq
	\lan\pm|i\hat{a}_1\hat{b}_L |\pm\ran = \pm 1 = \text{fermion parity} \label{eq:majorana-correlation-MF}
\eeq
in the ground states $|\pm\ran$~\cite{pfeuty,SS1,SS2,fu2010,vijay-fu,reslen2018end}. This
effect is sometimes called ``teleportation'' because there is a finite amplitude, in
the states $|\pm\ran$, to annihilate a fermion at one end of the wire and simultaneously create a 
fermion at the other end. This correlation is also responsible for the
$4\pi$-periodic Josephson effect~\cite{kitaev,kwon2004}, which is believed
to be a key physical signature of the topological superconducting phase. 

\begin{figure}[t]
  \centering
    \includegraphics[width= .35\textwidth]{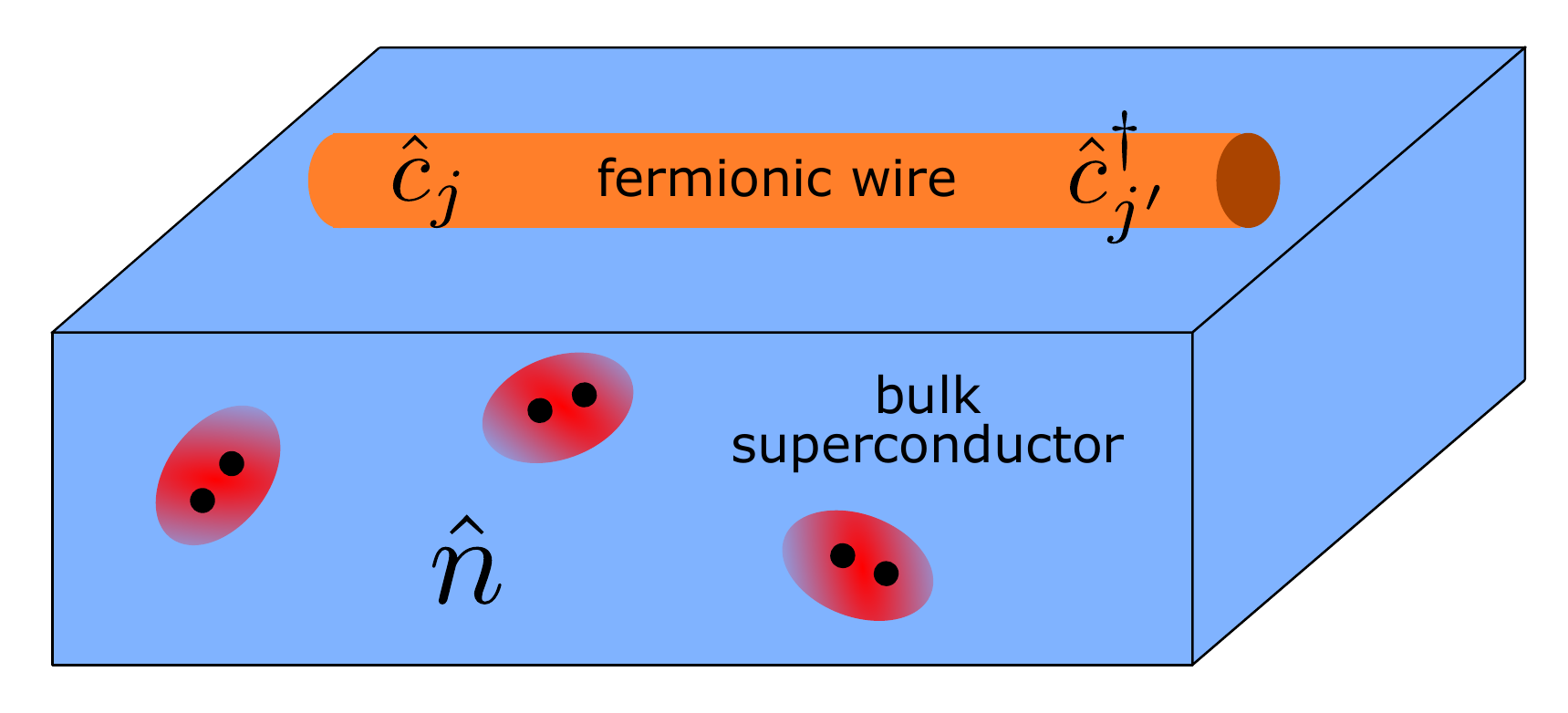} 
\caption{A fermionic wire on top of a bulk superconductor (SC). The operator $\hat{n}$ counts the
number of Cooper pairs in the bulk SC.}
\label{fig:TSC}
\end{figure}

\textit{A number-conserving model:}
We now introduce the number-conserving model studied in this 
paper. We again consider fermions $\ch_j$ on a wire of length $L$, and
we also introduce an additional degree of freedom to model a bulk 
superconductor (SC) proximity-coupled to the fermionic wire (see Fig.~\ref{fig:TSC}). 
This new degree of freedom is comprised of number and phase operators 
$\nh$ and $\ph$ that are canonically conjugate to each other,
$[\nh, e^{i\ph}]= e^{i\ph}$. Physically, $\nh$ counts the number of \emph{Cooper pairs} in the bulk SC,
and $e^{\pm i\ph}$ adds/removes a Cooper pair from the bulk SC. 
The Hilbert space $\mathcal{H}_{\text{SC}}$ of 
the number/phase degree of freedom is spanned by the states $|p\ran$, 
$p\in\mathbb{Z}$, obeying $\nh|p\ran = p |p\ran$ (and $e^{\pm i\ph}|p\ran = |p\pm 1\ran$). 

The conserved total particle number operator in our model is 
\beq
	\hat{N}= \hat{N}_w + 2\nh\ ,
\eeq
where $\hat{N}_w= \sum_{j=1}^L \ch^{\dg}_j \ch_j$ as before, and the factor of two is present 
because $\nh$ counts the number of Cooper \emph{pairs}. 
The Hamiltonian for our model consists of two terms:
\beq
	\hat{H}= \hat{H}_0 + \hat{V}\ .
\eeq
The first term, $\hat{H}_0$, is a number-conserving version of the mean-field Hamiltonian 
$\hat{H}_{\text{MF}}$, 
\begin{align}
	\hat{H}_0 =& -\frac{t}{2}\sum_{j=1}^{L-1}(\ch^{\dg}_j\ch_{j+1}+\text{h.c.}) - \frac{\Delta}{2}\sum_{j=1}^{L-1}(\ch_j\ch_{j+1}e^{i\hat{\phi}}+\text{h.c.}) \nnb \\
	-& \mu  \left(\hat{N}_w -\frac{L}{2}\right)\ .
\end{align}
We see that all pairing terms in $\hat{H}_0$ are now accompanied by a factor of $e^{\pm i\ph}$ to ensure 
commutation with $\hat{N}$ (when two fermions leave the wire they
become a Cooper pair in the SC, and vice-versa). The second term in $\hat{H}$ is a charging 
energy term for the bulk SC,
\beq
	\hat{V}= E_c(2\nh-2n_c)^2= 4 E_c (\nh-n_c)^2\ ,
\eeq 
where
\beq
	E_c=\frac{\mathcal{E}_c}{L}\ ,
\eeq
and $\mathcal{E}_c$ is a rescaled charging energy that we will hold fixed as $L \rightarrow \infty$. 
This scaling of $E_c$ with $L$ reflects the physical 
(Coulomb) scaling of the charging energy for a bulk SC with linear dimensions comparable
to that of the wire. The parameter $n_c$ is the preferred number of Cooper pairs in the bulk SC when it is 
isolated (i.e., not attached to the fermionic wire). We do not necessarily assume that $n_c$ is
an integer.

This model differs from previously studied models in two important ways. First,
the coupling to $e^{i\hat{\phi}}$ is non-local, unlike the coupling to local
bosons in Ref.~\onlinecite{cui2019}. This allows for the coexistence of the gap and 
Majorana-like correlations in our model. Second, our charging energy 
term involves the Cooper pair number operator $\hat{n}$, not the total number operator
$\hat{N}$ as in Refs.~\onlinecite{fu2010,alicea-milestones,scalable-designs}.
This allows us to study the effect of nontrivial superconducting phase fluctuations
on a TSC.

We now review some properties of this model at zero charging energy, $E_c=0$. 
As we show in the Supplemental Material~\footnote{See Supplemental Material for
additional results, the proofs of Claim 1, Lemma 1, Theorem 3, and Theorem 4, and our result on 
fermion parity switches. The Supplemental
Material includes Refs.~\onlinecite{horn,abbott,read2009,stone-chung,yang-ODLRO}.}, the eigenstates of 
$\hat{H}_0$ in the sector with 
fixed total particle number $\hat{N}=M$ are in one-to-one correspondence with the eigenstates
of $\hat{H}_{\text{MF}}$ in the sector with fermion parity $(-1)^{M}$, and the corresponding
energy eigenvalues are identical.

As in the mean-field case, $\hat{H}_0$ simplifies greatly at the special point $t=\Delta$ and
$\mu=0$. Let us denote $\hat{H}_0$ at this point by $\hat{H}_{0,*}$, and let
$|\chi^{(M)}_{0,*}\ran$ be the ground state of $\hat{H}_{0,*}$ in the
$M$-particle sector. One can show that $\hat{H}_{0,*}$
commutes with the pair of ``Majorana-like'' operators
$\hat{A}^{\dg}_1 = -i(e^{i\ph}\ch_1 -\ch^{\dg}_1)$ and 
$\hat{B}_L = \ch_L +e^{-i\ph}\ch^{\dg}_L$~\cite{scalable-designs}.
The operator 
$i\hat{A}_1^{\dg}\hat{B}_L$ is Hermitian and squares to the identity, and it
is the number-conserving analog of $i\hat{a}_1\hat{b}_L $ from the 
mean-field case. In addition, $|\chi^{(M)}_{0,*}\ran$ is an eigenstate of $i\hat{A}_1^{\dg}\hat{B}_L$ with
eigenvalue $(-1)^M$: 
i.e. $i\hat{A}_1^{\dg}\hat{B}_L|\chi^{(M)}_{0,*}\ran= (-1)^M |\chi^{(M)}_{0,*}\ran$. 
This implies the existence of a long range Majorana-like correlation,
\beq
	\lan \chi^{(M)}_{0,*}|i\hat{A}_1^{\dg}\hat{B}_L|\chi^{(M)}_{0,*}\ran= (-1)^{M}\ ,
	\label{eq:Majorana-correlation-star}
\eeq
which is the number-conserving analog of Eq.~\eqref{eq:majorana-correlation-MF}. This shows
that long range correlations between the ends of the wire also occur in the number-conserving model 
when $E_c=0$. One of our main results is that this correlation survives for $E_c\neq 0$. 

\textit{Results in a simple case:} We first present our results in the special case where
$t=\Delta$, $\mu=0$, and $M-2n_c= \frac{L}{2}$ (we work in the $M$-particle sector), as this case
illustrates the physical content of our results without introducing any 
unnecessary complications. 
We have already discussed the simplifications that occur at $t=\Delta$ and $\mu=0$.
The significance of the final condition $M-2n_c= \frac{L}{2}$ is that at this point the charging energy 
term $\hat{V}$ prefers the wire to have the same filling as in the ground state of 
$\hat{H}_{0,*}$, namely a filling equal to $\frac{L}{2}$. 
(We discuss this point in more detail below when we present our 
general results). We now state our results in this simple case. For technical reasons,
we assume here that $L$ is even and $L>2$.

The mean-field model has a unique ground state and a finite energy gap in a sector of fixed 
fermion parity. Our first theorem shows that, for small enough $\mathcal{E}_c$,
the number-conserving model has a unique ground state and a 
finite energy gap in a sector of fixed \emph{total particle number}.

\begin{thm}[energy gap]
If $\mathcal{E}_c < 4\Delta$, then
$\hat{H}$ has a unique ground state in the $M$-particle sector.
In addition, the energy gap $\delta^{(M)}$ of $\hat{H}$ in the $M$-particle sector is bounded from below as
\beq
	\delta^{(M)} \geq \Delta-\frac{\mathcal{E}_c}{4}\ ,
\eeq
and this bound holds for all system sizes $L$.
\end{thm}

Our second theorem shows that, for small enough $\mathcal{E}_c$,
the Majorana-like correlation of Eq.~\eqref{eq:Majorana-correlation-star} persists in the full
number-conserving model, in the sense that 
$(-1)^M\lan \psi^{(M)}_{0}|i\hat{A}_1^{\dg}\hat{B}_L|\psi^{(M)}_{0}\ran$ is bounded from below by a finite
positive constant as $L\to \infty$.

\begin{thm}[Majorana-like correlation]
Assume that the conditions of Theorem 1 are satisfied, and let $|\psi^{(M)}_0\ran$ be the unique ground 
state of $\hat{H}$ in the $M$-particle sector.  
Then 
\beq
	(-1)^M \lan \psi^{(M)}_{0}|i\hat{A}_1^{\dg}\hat{B}_L|\psi^{(M)}_{0}\ran \geq 1-\frac{\mathcal{E}_c}{2\Delta}\ .
\eeq
and this bound holds for all system sizes $L$. 
In particular, the function $1-\mathcal{E}_c/(2\Delta)$
is positive for $\mathcal{E}_c<2\Delta$.
\end{thm}

\textit{Proof of Theorem 1:} This proof uses the positivity of the charging energy term $\hat{V}$. 
Let $E^{(M)}_0 \leq E^{(M)}_1 \leq E^{(M)}_2 \leq \cdots$  and 
$\ep^{(M)}_0 \leq \ep^{(M)}_1 \leq \ep^{(M)}_2 \leq \cdots$ be the eigenvalues of
$\hat{H}$ and $\hat{H}_0$, respectively, within the $M$-particle sector.
Then the positivity of $\hat{V}$ implies that 
$\ep^{(M)}_a \leq E^{(M)}_a \ \forall\ a\in\{0,1,2,\dots\}$ (Corollary 4.3.12 in Ref.~\onlinecite{horn}).
In addition, if $|\chi^{(M)}_0\ran$ is the ground state of $\hat{H}_0$ in the
$M$-particle sector, then, by the variational theorem, we have
$E^{(M)}_0 \leq \ep^{(M)}_0 + \lan \chi^{(M)}_0|\hat{V}|\chi^{(M)}_0\ran$. By
combining this result with the bounds implied by the positivity of $\hat{V}$, we 
find that $E^{(M)}_1 - E^{(M)}_0 \geq \ep^{(M)}_1 - \ep^{(M)}_0 - 
\lan \chi^{(M)}_0|\hat{V}|\chi^{(M)}_0\ran$, or
\beq
	\delta^{(M)} \geq \delta^{(M)}_0 - \lan \chi^{(M)}_0|\hat{V}|\chi^{(M)}_0\ran\ ,
\eeq
where $\delta^{(M)}= E^{(M)}_1 - E^{(M)}_0$ and $\delta^{(M)}_0= \ep^{(M)}_1 - \ep^{(M)}_0$ are
the energy gaps (in the $M$-particle sector) of $\hat{H}$ and $\hat{H}_0$, respectively.
This inequality holds for general parameter values $t,\Delta,\mu,\mathcal{E}_c$, and $n_c$, and it
implies that $\hat{H}$ has a unique ground state in the
$M$-particle sector if $\delta^{(M)}_0 > \lan \chi^{(M)}_0|\hat{V}|\chi^{(M)}_0\ran$.

To complete the proof we need to compute 
$\lan \chi^{(M)}_0|\hat{V}|\chi^{(M)}_0\ran$. At this point we specialize to 
$t=\Delta$, $\mu=0$, and $M-2n_c=\frac{L}{2}$, and so the state $|\chi^{(M)}_0\ran=|\chi^{(M)}_{0,*}\ran$
is the ground state of $\hat{H}_{0,*}$. We
can compute $\lan \chi^{(M)}_{0,*}|\hat{V}|\chi^{(M)}_{0,*}\ran$ exactly using 
Eq.~4.10 in the Supplemental Material, and we find that 
\beqa
	\lan \chi^{(M)}_{0,*}|\hat{V}|\chi^{(M)}_{0,*}\ran &=&  E_c\lan \chi^{(M)}_{0,*}|\left(\hat{N}_w -\frac{L}{2}\right)^2|\chi^{(M)}_{0,*}\ran \nnb \\
	&=& \frac{\mathcal{E}_c}{4}\ ,\ M=\text{even or odd}\ .
\eeqa
Finally, using the fact that $\delta^{(M)}_0=\Delta$ for $t=\Delta$ and $\mu=0$, we arrive at the
result of Theorem 1.

\textit{Proof of Theorem 2:}
The proof of Theorem 2 is based on a result (Claim 1) that bounds the overlap of the 
ground state $|\psi^{(M)}_0\ran$ of $\hat{H}$ with the ground state $|\chi^{(M)}_0\ran$ of $\hat{H}_0$. 
We prove Claim 1 in the Supplemental Material. 

\begin{claim}
Assume that $\delta^{(M)}= E^{(M)}_1 - E^{(M)}_0$ is positive: 
$\delta^{(M)}>0$. Then 
\begin{align}
	|\lan \chi^{(M)}_0| \psi^{(M)}_0\ran|^2 \geq 1 - \frac{\lan \chi^{(M)}_0 |\hat{V}|\chi^{(M)}_0\ran}{\delta^{(M)}_0}\ ,
\end{align}
and this result holds for general values of the parameters $t,\Delta,\mu,\mathcal{E}_c,$ and $n_c$.
\end{claim}

We now use Claim 1 to prove Theorem 2.
We again specialize to 
$t=\Delta$, $\mu=0$, and $M-2n_c=\frac{L}{2}$, and so $|\chi^{(M)}_0\ran$ is again 
the ground state $|\chi^{(M)}_{0,*}\ran$ of $\hat{H}_{0,*}$. We also define the projectors
$\hat{P}_{\chi}= |\chi^{(M)}_{0,*}\ran \lan \chi^{(M)}_{0,*}|$ and $\hat{Q}_{\chi}=1-\hat{P}_{\chi}$, 
and we note that $[\hat{P}_{\chi},i\hat{A}_1^{\dg}\hat{B}_L]=0$.

We first prove the theorem for even $M$. Using 
$i\hat{A}_1^{\dg}\hat{B}_L|\chi^{(M)}_{0,*}\ran= (-1)^M |\chi^{(M)}_{0,*}\ran$ and
$[\hat{Q}_{\chi},i\hat{A}_1^{\dg}\hat{B}_L]=0$, we have
\begin{align}
	\lan &\psi^{(M)}_{0}|i\hat{A}_1^{\dg}\hat{B}_L|\psi^{(M)}_{0}\ran \nnb \\
	&= \lan \psi^{(M)}_{0}|
	i\hat{A}_1^{\dg}\hat{B}_L\hat{P}_{\chi}|\psi^{(M)}_{0}\ran + \lan \psi^{(M)}_{0}|i\hat{A}_1^{\dg}\hat{B}_L\hat{Q}_{\chi}|\psi^{(M)}_{0}\ran  \nnb \\
	&=  |\lan \chi^{(M)}_{0,*}| \psi^{(M)}_{0}\ran|^2 + \lan \psi^{(M)}_{0}|\hat{Q}_{\chi}i\hat{A}_1^{\dg}\hat{B}_L\hat{Q}_{\chi}|\psi^{(M)}_{0}\ran \nnb \\
	&\geq |\lan \chi^{(M)}_{0,*}| \psi^{(M)}_{0}\ran|^2 - \lan \psi^{(M)}_{0}|\hat{Q}_{\chi}^2|\psi^{(M)}_{0}\ran \ ,
\end{align}
where we used the fact that 
$\lan \vphi|i\hat{A}_1^{\dg}\hat{B}_L|\vphi\ran \geq -\lan\vphi|\vphi\ran$ for any
state $|\vphi\ran$ (since the eigenvalues of
$i\hat{A}_1^{\dg}\hat{B}_L$ are $1$ and $-1$). 
The last line can be rewritten in the form
\beq
	\lan \psi^{(M)}_{0}|i\hat{A}_1^{\dg}\hat{B}_L|\psi^{(M)}_{0}\ran \geq  2|\lan \chi^{(M)}_{0,*}| \psi^{(M)}_{0}\ran|^2-1\ .
\eeq
For odd $M$ we again use 
$i\hat{A}_1^{\dg}\hat{B}_L|\chi^{(M)}_{0,*}\ran= (-1)^M |\chi^{(M)}_{0,*}\ran$, but now we
use $\lan \vphi|i\hat{A}_1^{\dg}\hat{B}_L|\vphi\ran \leq \lan\vphi|\vphi\ran$ for any
state $|\vphi\ran$. We then find that
\beq
		\lan \psi^{(M)}_{0}|i\hat{A}_1^{\dg}\hat{B}_L|\psi^{(M)}_{0}\ran \leq  -2|\lan \chi^{(M)}_{0,*}| \psi^{(M)}_{0}\ran|^2+1\ .
\eeq
We now use Claim 1 to replace 
$|\lan \chi^{(M)}_{0,*}| \psi^{(M)}_{0}\ran|^2$ in these inequalities with its lower bound. 
The result for either parity of $M$ can be expressed as
\beq
	(-1)^M\lan \psi^{(M)}_{0}|i\hat{A}_1^{\dg}\hat{B}_L|\psi^{(M)}_{0}\ran \geq 1 - 2\frac{\lan \chi^{(M)}_{0,*} |\hat{V}|\chi^{(M)}_{0,*}\ran}{\delta^{(M)}_0}\ .
\eeq
Finally, we use $\lan \chi^{(M)}_{0,*} |\hat{V}|\chi^{(M)}_{0,*}\ran=\mathcal{E}_c/4$ and  
$\delta^{(M)}_0=\Delta$ for $t=\Delta$ and $\mu=0$ to arrive at the result of Theorem 2.

\textit{Results in the general case:} We now present our results in the
general case. To start, we rewrite the Hamiltonian as 
$\hat{H}_{\mu}= \hat{H}_{0,\mu}+\hat{V}$ to emphasize its dependence on the wire
chemical potential $\mu$. Similarly, we rewrite the mean-field Hamiltonian with chemical potential $\mu$ 
as $\hat{H}_{\text{MF},\mu}$. We also define $|\chi^{(M)}_{0,\mu}\ran$ and $\delta^{(M)}_{0,\mu}$ 
to be the ground state and energy gap of $\hat{H}_{0,\mu}$ in the $M$-particle sector 
(note that $\delta^{(M)}_{0,\mu}$ is equal to the gap of 
$\hat{H}_{\text{MF},\mu}$ in the sector with fermion parity $(-1)^M$). 
In what follows, we assume that $|\mu|<t$ and $\Delta\neq 0$ (the criteria
for the mean-field topological phase).

Next, we define the function $f^{(M)}(\mu)$ to be equal to the average fermion number
on the wire in the state $|\chi^{(M)}_{0,\mu}\ran$, 
\beq
	f^{(M)}(\mu)= \lan \chi^{(M)}_{0,\mu}|\hat{N}_w|\chi^{(M)}_{0,\mu}\ran\ .
\eeq
The results in the Supplemental Material imply that $f^{(M)}(\mu)$ is also equal to the expectation 
value of $\hat{N}_w$ in the ground state of $\hat{H}_{\text{MF},\mu}$ in the sector 
with fermion parity $(-1)^M$. The charging energy term, when restricted to the $M$-particle sector, 
usually prefers the wire to have a different filling than $f^{(M)}(\mu)$. 
To see this we note that, \emph{within the $M$-particle sector}, $\hat{V}$ can be rewritten 
as
\beq
	\hat{V}= 4 E_c(\nh-n_c)^2 = E_c\left[\hat{N}_w - (M - 2n_c)\right]^2 \ .
\eeq
We see that this term prefers the filling of the wire to be $M-2n_c$, which in general is \emph{not} 
equal to $f^{(M)}(\mu)$. This will be important in what follows. 

Finally, we need to discuss the analog of the Majorana-like operators $\hat{A}_1$ and 
$\hat{B}_L$ in the general case. For any $t,\Delta,\mu$ such
that $|\mu|<t$ and $\Delta\neq 0$, one can construct operators 
$\hat{\Gamma}_{\ell,\mu}$ and $\hat{\Gamma}_{r,\mu}$ ($\ell/r$ = left/right) such that (i) 
$\hat{\Gamma}_{\ell,\mu}$ and $\hat{\Gamma}_{r,\mu}$ obey the same algebra as $\hat{A}_1$ and $\hat{B}_L$, 
(ii) $\hat{\Gamma}_{\ell/r,\mu}$ is localized near the left/right end of the wire, and (iii) 
$|\chi^{(M)}_{0,\mu}\ran$ is an eigenvector of $i\hat{\Gamma}_{\ell,\mu}^{\dg}\hat{\Gamma}_{r,\mu}$ 
with eigenvalue equal to $(-1)^M$: i.e. 
$i\hat{\Gamma}_{\ell,\mu}^{\dg}\hat{\Gamma}_{r,\mu}|\chi^{(M)}_{0,\mu}\ran= 
(-1)^M |\chi^{(M)}_{0,\mu}\ran$. 
These properties imply that $|\chi^{(M)}_{0,\mu}\ran$ also possesses a Majorana-like 
correlation 
\beq
	\lan \chi^{(M)}_{0,\mu}|i\hat{\Gamma}_{\ell,\mu}^{\dg}\hat{\Gamma}_{r,\mu}|\chi^{(M)}_{0,\mu}\ran= (-1)^{M}\ .
	\label{eq:Majorana-correlation}
\eeq
The operators $\hat{\Gamma}_{\ell,\mu}$ and $\hat{\Gamma}_{r,\mu}$ can be constructed from  
$\hat{A}_1$ and $\hat{B}_L$ using the exact version~\cite{Osborne} of 
\emph{quasiadiabatic continuation}~\cite{QAC}, as discussed in a related 
context in Sec.~VII of Ref.~\onlinecite{stabilityTO}. This construction uses the fact that,
for $|\mu|<t$ and $\Delta\neq 0$, $\hat{H}_{\text{MF},\mu}$ can be deformed to the special point $t = \Delta$ and $\mu = 0$ without closing the energy gap of the system~\footnote{In more detail: the first step of the construction is to apply quasiadiabatic continuation to the mean-field model 
$\hat{H}_{\text{MF},\mu}$ to construct operators 
$\hat{\gamma}_{\ell,\mu}$ and $\hat{\gamma}_{r,\mu}$ such that (i) 
$\hat{\gamma}_{\ell,\mu}$ and $\hat{\gamma}_{r,\mu}$ obey the same algebra as $\hat{a}_1$ and $\hat{b}_L$, 
(ii) $\hat{\gamma}_{\ell/r,\mu}$ is localized near the left/right end of the wire, and (iii) the
ground state of $\hat{H}_{\text{MF},\mu}$ in a sector of fixed fermion parity is an eigenvector of 
$i\hat{\gamma}_{\ell,\mu}\hat{\gamma}_{r,\mu}$ with eigenvalue equal to its fermion parity.
The operators $\hat{\Gamma}_{\ell/r,\mu}$ for the number-conserving model $\hat{H}_{0,\mu}$ are then
obtained from $\hat{\gamma}_{\ell/r,\mu}$ by making the replacements $\ch_j \to \ch_j$ and
$\ch^{\dg}_j \to \ch^{\dg}_j e^{-i\ph}$.}. 
The only difference between $\hat{\Gamma}_{\ell,\mu}$, $\hat{\Gamma}_{r,\mu}$ and 
$\hat{A}_1$, $\hat{B}_L$ is that the latter are localized exactly at the ends of the wire, 
while the former extend into the bulk of the wire (but they decay rapidly due to the 
locality-preserving property of the quasiadiabatic continuation technique). 

We now present a lemma on which our later results are built. This lemma allows us to rewrite the
Hamiltonian in a way that removes the tension between the wire chemical potential
$\mu$ and the filling $M-2n_c$ preferred by the charging energy term.

\begin{lemma}
Consider the $M$-particle sector, and assume that
\beq
	-\frac{(t+\mu)}{2\mathcal{E}_c}L < M-2n_c - f^{(M)}(\mu) < \frac{(t-\mu)}{2\mathcal{E}_c}L\ , \label{eq:interval}
\eeq
Then there is a unique chemical potential $\mu'$ satisfying $|\mu'| < t$
and such that (within the $M$-particle sector) 
\beq
	\hat{H}_{\mu}= \hat{H}_{0,\mu'} + E_c \left[\hat{N}_w - f^{(M)}(\mu')\right]^2 + \mathcal{C}\ ,
\eeq
where $\mathcal{C}$ is an overall constant. 
\end{lemma}

This lemma shows that, if Eq.~\eqref{eq:interval} holds, then $\hat{H}_{\mu}$ can be 
rewritten in such a
way that the charging energy term now prefers the same filling $f^{(M)}(\mu')$ of the wire as in
the ground state $|\chi^{(M)}_{0,\mu'}\ran$ of the new unperturbed Hamiltonian $\hat{H}_{0,\mu'}$.
In addition, since $\mu'$ also satisfies $|\mu'|<t$, a mean-field Hamiltonian
with parameters $t,\Delta,\mu'$ is also in the mean-field 
topological phase. The new chemical potential $\mu'$ is in general
a function of all of the constants $t,\Delta,\mu,\mathcal{E}_c,n_c,M$, and $L$.
Finally, we note that the size of the interval in 
Eq.~\eqref{eq:interval} is proportional to $L$, which means that the result of Lemma 1 applies
for a large range of parameter values.

We can now state our main results in full generality. These
results are generalizations of Theorems 1 and 2 to the case of general
$t,\Delta,\mu,\mathcal{E}_c$, and $n_c$. For technical reasons, we
assume here that $L$ is even.

\begin{thm}[energy gap]
Assume that Eq.~\eqref{eq:interval} of Lemma 1 is 
satisfied, and let $\mu'$ be the new chemical potential guaranteed by Lemma 1.
If $\mathcal{E}_c\left(\frac{1}{2}+\frac{8}{L}\right) < \delta^{(M)}_{0,\mu'}$, then
$\hat{H}_{\mu}$ has a unique ground state in the $M$-particle sector.
In addition, the energy gap $\delta^{(M)}_{\mu}$ of $\hat{H}_{\mu}$ in the $M$-particle sector
is bounded from below as
\beq
	\delta^{(M)}_{\mu} \geq \delta^{(M)}_{0,\mu'}-\mathcal{E}_c\left(\frac{1}{2}+\frac{8}{L}\right) = \delta^{(M)}_{0,\mu'}-\frac{\mathcal{E}_c}{2} + O\left(\frac{1}{L}\right)\ ,
\eeq
and this bound holds for all system sizes $L$. 
\end{thm}

\begin{thm}[Majorana-like correlation]
Assume that Eq.~\eqref{eq:interval} of 
Lemma 1 is satisfied, and let $\mu'$ be the new chemical potential guaranteed by Lemma 1. 
Also, assume that the conditions of Theorem 3 are satisfied, and 
let $|\psi^{(M)}_{0,\mu}\ran$ be the unique ground state of $\hat{H}_{\mu}$ in
the $M$-particle sector. 
Then 
\begin{align}
	(-1)^M\lan \psi^{(M)}_{0,\mu}|i\hat{\Gamma}_{\ell,\mu'}^{\dg}\hat{\Gamma}_{r,\mu'}|\psi^{(M)}_{0,\mu}\ran \geq 1-\frac{\mathcal{E}_c}{\delta^{(M)}_{0,\mu'}} + O\left(\frac{1}{L}\right) \ .
\end{align}
In particular, the function $1-\mathcal{E}_c/\delta^{(M)}_{0,\mu'}$
is positive for $\mathcal{E}_c<\delta^{(M)}_{0,\mu'}$.
\end{thm}

It is important to note that in these theorems, the Majorana-like operators $\hat{\Gamma}_{\ell/r,\mu'}$ 
and unperturbed energy gap $\delta^{(M)}_{0,\mu'}$ are those for the 
Hamiltonian $\hat{H}_{0,\mu'}$ with the new chemical potential $\mu'$, and not those for
the original chemical potential $\mu$. 

We present the proofs of Lemma 1, Theorem 3, and Theorem 4 in the Supplemental Material.

\textit{Fermion parity switching:} In the Supplemental Material we prove a theorem (Theorem 5) that
shows that for small enough $\mathcal{E}_c$, 
the lowest energy state of our model 
in a ring geometry with periodic boundary conditions occurs in a sector with 
\emph{odd} total particle number, while the lowest energy state with anti-periodic boundary conditions
occurs in a sector with \emph{even} total particle number.
This shows that our model displays a ``fermion parity switch'', just like the Kitaev p-wave 
wire model~\cite{kitaev}. We also note that Refs.~\onlinecite{ortiz1,ortiz2} proved that a fermion 
parity switch occurs in their exactly solvable model of a number-conserving TSC.

\textit{Discussion:} We have proved that many of the properties of a single 
mean-field topological superconducting wire are also present in a more realistic number-conserving model.
Because of the non-locality present in our model, we could not apply the methods of 
proof from Ref.~\onlinecite{stabilityTO}. 
Instead, our proofs rely on the positivity of the charging energy term 
$\hat{V}$, and on the small magnitude ($\propto \sqrt{L}$) of particle-number fluctuations in 1D 
mean-field models of superconductivity. Our methods can also be used to prove similar results in 
models with \emph{spinful} fermions, for example number-conserving
versions of the experimentally relevant models of 
Refs.~\onlinecite{PhysRevLett.105.077001,PhysRevLett.105.177002}. An important challenge for the
future is to investigate the stability
of proposed architectures for Majorana-based qubits~\cite{alicea-milestones,scalable-designs} in the 
more realistic number-conserving setting considered here. 
The crucial open question is whether the finite-size energy splitting of 
qubit states remains exponentially small in the system size $L$.

\textit{Acknowledgements:} We thank G. Ortiz for a discussion about fermion parity
switches. M.F.L. and M.L. 
acknowledge the support of the Kadanoff Center for Theoretical Physics at the University of Chicago.
This work was supported by the Simons Collaboration on Ultra-Quantum Matter, which is a grant from the 
Simons Foundation (651440, M.L.).


%

\end{document}


\title{Supplemental Material for ``Rigorous results on topological superconductivity with particle number conservation''}

\author{Matthew F. Lapa}
\email[email address: ]{mlapa@uchicago.edu}
\affiliation{Kadanoff Center for Theoretical Physics, University of Chicago, Chicago, IL, 60637, USA}

\author{Michael Levin}
\email[email address: ]{malevin@uchicago.edu}
\affiliation{Kadanoff Center for Theoretical Physics, University of Chicago, Chicago, IL, 60637, USA}

\maketitle

\section{Proofs of Claim 1, Lemma 1, Theorem 3, and Theorem 4}

\subsection{Proof of Claim 1}

In this section we present the proof of Claim 1. Let 
$\hat{P}_{\psi}= |\psi^{(M)}_0\ran \lan \psi^{(M)}_0|$ 
be the projector onto the ground state of $\hat{H}$, and let 
$\hat{Q}_{\psi}=1-\hat{P}_{\psi}$. Then we have
\beqa
	\lan\chi^{(M)}_0|\hat{H}|\chi^{(M)}_0\ran &=& \lan\chi^{(M)}_0|\hat{P}_{\psi}\hat{H}\hat{P}_{\psi}|\chi^{(M)}_0\ran + \lan\chi^{(M)}_0|\hat{Q}_{\psi}\hat{H}\hat{Q}_{\psi}|\chi^{(M)}_0\ran \nnb \\
	&=& E^{(M)}_0|\lan\chi^{(M)}_0|\psi^{(M)}_0\ran|^2 + \lan\chi^{(M)}_0|\hat{Q}_{\psi}\hat{H}\hat{Q}_{\psi}|\chi^{(M)}_0\ran \nnb \\
	&\geq&  E^{(M)}_0|\lan\chi^{(M)}_0|\psi^{(M)}_0\ran|^2 + E^{(M)}_1 \lan\chi^{(M)}_0|\hat{Q}_{\psi}^2|\chi^{(M)}_0\ran\ ,
\eeqa
where $E^{(M)}_0$ and $E^{(M)}_1\geq E^{(M)}_0$ are the two lowest eigenvalues of $\hat{H}$ in the 
$M$-particle sector. This last inequality can be rewritten as
\beqa
	\lan\chi^{(M)}_0|\hat{H}|\chi^{(M)}_0\ran \geq E^{(M)}_1 - (E^{(M)}_1 - E^{(M)}_0) |\lan\chi^{(M)}_0|\psi^{(M)}_0\ran|^2\ ,
\eeqa
and then we can use $E^{(M)}_0 \geq \ep^{(M)}_0$ (which follows from the
positivity of $\hat{V}$) to obtain
\beq
	\lan\chi^{(M)}_0|\hat{H}|\chi^{(M)}_0\ran \geq E^{(M)}_1 - (E^{(M)}_1 - \ep^{(M)}_0) |\lan\chi^{(M)}_0|\psi^{(M)}_0\ran|^2\ .
\eeq
Next, using our assumption that $\delta^{(M)}=E^{(M)}_1 - E^{(M)}_0 >0$ (which implies that
$E^{(M)}_1 - \ep^{(M)}_0>0$ since $E^{(M)}_0 \geq \ep^{(M)}_0$), 
we can rearrange this expression to obtain
\beq
	|\lan\chi^{(M)}_0|\psi^{(M)}_0\ran|^2 \geq  \frac{E^{(M)}_1 - \lan\chi^{(M)}_0|\hat{H}|\chi^{(M)}_0\ran }{E^{(M)}_1 - \ep^{(M)}_0}\ .
\eeq

To proceed further we first note that $\lan\chi^{(M)}_0|\hat{H}|\chi^{(M)}_0\ran \geq \ep^{(M)}_0$,
which also follows from the positivity of $\hat{V}$. Next, we note that, for
real numbers $a$ and $b$ satisfying $a \geq b$, the function $f(z)= \frac{z-a}{z-b}$ is an \emph{increasing}
function of $z$ for $z>b$ (we restrict our attention to values of $z$ to the right of the
pole at $z=b$). We now apply this to our last inequality by choosing 
$a= \lan\chi^{(M)}_0|\hat{H}|\chi^{(M)}_0\ran$ and $b= \ep^{(M)}_0$, and
using $E^{(M)}_1\geq \ep^{(M)}_1$ (which again follows from the positivity of 
$\hat{V}$). We find that
\beq
	\frac{E^{(M)}_1 - \lan\chi^{(M)}_0|\hat{H}|\chi^{(M)}_0\ran }{E^{(M)}_1 - \ep^{(M)}_0} \geq \frac{\ep^{(M)}_1 - \lan\chi^{(M)}_0|\hat{H}|\chi^{(M)}_0\ran }{\ep^{(M)}_1 - \ep^{(M)}_0}\ ,
\eeq
which leads to
\beq
	|\lan\chi^{(M)}_0|\psi^{(M)}_0\ran|^2 \geq 1 - \frac{\lan\chi^{(M)}_0|\hat{V}|\chi^{(M)}_0\ran}{\delta^{(M)}_0}\ ,
\eeq
where $\delta^{(M)}_0=  \ep^{(M)}_1 - \ep^{(M)}_0$ is the energy gap of $\hat{H}_0$ in the M-particle
sector. This completes the proof of Claim 1.

\subsection{Proof of Lemma 1}

In this section we present the proof of Lemma 1. We start
by rewriting $\hat{H}_{\mu}$ in the desired final form 
\beq
	\hat{H}_{\mu}= \hat{H}_{0,\mu'} + E_c \left[\hat{N}_w - f^{(M)}(\mu')\right]^2 + \mathcal{C}\ .
\eeq
[Recall that we work within the $M$-particle sector, where $\hat{N}_w + 2\hat{n}= M$.]
By comparing with the original expression for $\hat{H}_{\mu}$, 
we see that this expression can only be valid if $\mu'$ satisfies the equation
\beq
	\mu + 2E_c (M-2n_c)= \mu' + 2E_c f^{(M)}(\mu')\ . \label{eq:mu-prime-eqn}
\eeq
If this equation can be satisfied, then $\hat{H}_{\mu}$ can be rewritten as shown above with the
constant $\mathcal{C}$ given by 
\beq
	\mathcal{C}= (\mu-\mu')\frac{L}{2}+E_c (M-2n_c)^2-E_c [f^{(M)}(\mu')]^2  \ .
\eeq

Our goal, then, is to prove the existence of a $\mu'$ satisfying Eq.~\eqref{eq:mu-prime-eqn}. 
Recall that we assume that $|\mu|<t$, and that we also want $|\mu'|<t$ (so that both
Hamiltonians $\hat{H}_{\text{MF},\mu}$ and $\hat{H}_{\text{MF},\mu'}$ are in the
mean-field topological phase). We now prove that, if $|\mu| < t$ and if 
Eq.~21 of the main text holds, then there is a unique $\mu'$ satisfying $|\mu'|<t$ and 
Eq.~\eqref{eq:mu-prime-eqn}. This $\mu'$ will be a function of all of the parameters 
$t,\Delta,\mu,\mathcal{E}_c,n_c,M$, and $L$ present in the model.

The strategy of the proof is as follows. First, note that $|\mu|<t$ implies that
$|\mu|<t-\omega$ for any $\omega\in(0,t-|\mu|)$. Now consider the 
function 
\beq
 	g^{(M)}(x)=x + 2E_c f^{(M)}(x)\ .
\eeq 
This function, evaluated at the point $x=\mu'$, is what appears on the right-hand side of 
Eq.~\eqref{eq:mu-prime-eqn}. We now prove several important facts about this function. First, we
prove that $g^{(M)}(x)$ is a \emph{strictly increasing} function of $x$.
Next, we show that $g^{(M)}(x)$ has the \emph{intermediate value property} for closed intervals 
$[a,b]\subset (-t,t)$, namely, for $-t<a<b<t$, if $g^{(M)}(a)< y <g^{(M)}(b)$ (or
$g^{(M)}(b)< y <g^{(M)}(a)$), then there exists
$c\in(a,b)$ such that $g^{(M)}(c)= y$. Next, for any $\omega\in(0,t-|\mu|)$
we establish a lower bound on $g^{(M)}(x)$
at $x=t-\omega$, $g^{(M)}(t-\omega) \geq \eta_U(\omega)$, and an upper bound on $g^{(M)}(x)$
at $x=-t+\omega$, $g^{(M)}(-t+\omega) \leq \eta_L(\omega)$. [The subscripts $U/L$ stand for upper/lower.]
In this situation the intermediate value property for $g^{(M)}(x)$ implies that if
\beq
	\eta_L(\omega) < \mu + 2E_c (M-2n_c) < \eta_U(\omega)\ ,
\eeq
then there exists a $\mu'\in(-t+\omega,t-\omega)$ such that Eq.~\eqref{eq:mu-prime-eqn} holds.
In addition, this $\mu'$ is actually unique since $g^{(M)}(x)$ is a strictly
increasing function of $x$. Finally, we complete the proof by using this result,
which holds for \emph{any} $\omega\in(0,t-|\mu|)$, to prove a stronger result in which we can take 
$\omega=0$ in all formulas.

We start by proving that $f^{(M)}(x)$ is an \emph{increasing} function of $x$, i.e., that 
$x_2 > x_1 \Rightarrow f^{(M)}(x_2) \geq f^{(M)}(x_1)$. This is enough to show that
$g^{(M)}(x)$ is a strictly increasing function of $x$, since the linear function 
$x$ is strictly increasing. That $f^{(M)}(x)$ is an increasing function of $x$ can be proven
using a variational argument. As before, let $|\chi^{(M)}_{0,x}\ran$ be the ground state of 
$\hat{H}_{0,x}$ in the $M$-particle sector, and let $\ep^{(M)}_{0,x}$ be the corresponding energy 
eigenvalue, $\hat{H}_{0,x}|\chi^{(M)}_{0,x}\ran= \ep^{(M)}_{0,x}|\chi^{(M)}_{0,x}\ran$. For 
a given $x_1$ and $x_2$, we can clearly write
\beq
	\hat{H}_{0,x_2}= \hat{H}_{0,x_1} + (x_1-x_2)\left(\hat{N}_w-\frac{L}{2}\right)\ ,
\eeq
and a similar equation holds if we swap $x_1$ and $x_2$. 
Using $|\chi^{(M)}_{0,x_1}\ran$ as a trial ground state for $\hat{H}_{0,x_2}$, we 
obtain
\beq
	\ep^{(M)}_{0,x_2}\leq \ep^{(M)}_{0,x_1} + (x_1-x_2)f^{(M)}(x_1)-(x_1-x_2)\frac{L}{2}\ .
\eeq
Similarly, using $|\chi^{(M)}_{0,x_2}\ran$ as a trial ground state for $\hat{H}_{0,x_1}$, we 
obtain
\beq
	\ep^{(M)}_{0,x_1}\leq \ep^{(M)}_{0,x_2} + (x_2-x_1)f^{(M)}(x_2)-(x_2-x_1)\frac{L}{2}\ .
\eeq
If we use this second inequality to bound $\ep^{(M)}_{0,x_1}$ on the right-hand side of the first inequality, 
then we find that
\beq
	\ep^{(M)}_{0,x_2} \leq \ep^{(M)}_{0,x_2} + (x_2-x_1)f^{(M)}(x_2)-(x_2-x_1)\frac{L}{2} + (x_1-x_2)f^{(M)}(x_1)-(x_1-x_2)\frac{L}{2}\ .
\eeq
This last inequality can be rewritten in the form
\beq
	0\leq (x_2-x_1)\left[f^{(M)}(x_2)-f^{(M)}(x_1) \right]\ ,
\eeq
which yields the result that $x_2 > x_1 \Rightarrow f^{(M)}(x_2) \geq f^{(M)}(x_1)$. 
As we mentioned above, this result implies that $g^{(M)}(x)$ is a strictly increasing 
function of $x$. 

We now prove that $g^{(M)}(x)$ obeys the intermediate value property for closed intervals 
$[a,b]\subset(-t,t)$. To prove this we first consider
some $x\in(-t,t)$, and some deviation $\delta x$, and write
$\hat{H}_{0,x+\delta x}= \hat{H}_{0,x} - \delta x\left(\hat{N}_w - \frac{L}{2}\right)$. Then, since
$\hat{H}_{0,x}$ has a unique ground state and an energy gap within the $M$-particle sector, Theorem 
6.3.12 of Ref.~\onlinecite{horn} implies that the ground state energy $\ep^{(M)}_{0,x+\delta x}$ of  
$\hat{H}_{0,x+\delta x}$ (in the $M$-particle sector)
is differentiable (with respect to $\delta x$) at $\delta x = 0$. In other words, $\ep^{(M)}_{0,x}$ is 
a differentiable function of $x$ for all $x\in(-t,t)$. In addition, the same 
theorem tells us that $\frac{d}{dx}\ep^{(M)}_{0,x}=L/2-f^{(M)}(x)$ (this is equivalent to the
Hellmann-Feynman theorem). Since $\ep^{(M)}_{0,x}$ is differentiable for $x\in(-t,t)$, Darboux's theorem 
(Theorem 5.2.7 in Ref.~\onlinecite{abbott})
then tells us that $\frac{d}{dx}\ep^{(M)}_{0,x}$ obeys the intermediate value property for any
closed interval $[a,b]$ contained within $(-t,t)$. Finally, since 
$g^{(M)}(x)= x + 2 E_c \left( L/2-\frac{d}{dx}\ep^{(M)}_{0,x} \right)$, it follows that 
$g^{(M)}(x)$ also obeys the intermediate value property (this is because 
$g^{(M)}(x)$ is equal to a continuous function of $x$ minus a constant times $\frac{d}{dx}\ep^{(M)}_{0,x}$).

We now use the increasing property of $f^{(M)}(x)$ 
to obtain the desired bounds $\eta_U(\omega)$ and $\eta_L(\omega)$ mentioned above. 
Consider the value of the function $g^{(M)}(x)$ at $x=\pm (t-\omega)$. Since 
$|\mu| < t-\omega$, we have
\beqa
	g^{(M)}(t-\omega) &=& t-\omega + 2 E_c f^{(M)}(t-\omega) \nnb \\
	&\geq& t-\omega + 2 E_c f^{(M)}(\mu) \equiv \eta_U(\omega)\ .
\eeqa
Similarly, we have
\beq
	g^{(M)}(-t+\omega) \leq -t+\omega + 2 E_c f^{(M)}(\mu) \equiv \eta_L(\omega)\ .
\eeq 
Using these bounds we find that, if
\beq
	-t+\omega + 2 E_c f^{(M)}(\mu)< \mu + 2E_c (M-2n_c) < t-\omega + 2 E_c f^{(M)}(\mu)\ , \label{eq:condition-omega}
\eeq
then there exists a unique $\mu'\in(-t+\omega,t-\omega)$ that satisfies Eq.~\eqref{eq:mu-prime-eqn}.  

We now use this result to prove the stronger result that if
\beq
	-t + 2 E_c f^{(M)}(\mu)< \mu + 2E_c (M-2n_c) < t + 2 E_c f^{(M)}(\mu)\ , \label{eq:condition-no-omega}
\eeq
then there exists a unique $\mu'\in(-t,t)$ that satisfies Eq.~\eqref{eq:mu-prime-eqn}. We
first prove existence of $\mu'\in(-t,t)$ under the new condition \eqref{eq:condition-no-omega}, and
then we prove uniqueness using an argument by contradiction. To prove existence we note that, by
the properties of open intervals, Eq.~\eqref{eq:condition-no-omega} implies that
Eq.~\eqref{eq:condition-omega} holds for some $\omega>0$, and we can 
always choose this $\omega$ small enough so that $|\mu|<t-\omega$ as well. Then our previous
results show that there exists a $\mu'\in(-t,t)$ satisfying Eq.~\eqref{eq:mu-prime-eqn}. 
[This is because our previous results showed the existence of such a $\mu'$ in the
interval $(-t+\omega,t-\omega)$, which is contained within the interval $(-t,t)$.]
To prove uniqueness we suppose, by contradiction, that Eq.~\eqref{eq:condition-no-omega} holds and
that there are two chemical potentials $\mu'_1,\mu'_2\in(-t,t)$ that both
satisfy Eq.~\eqref{eq:mu-prime-eqn}. Then $\mu,\mu'_1$, $\mu'_2$, and 
$\mu + 2E_c (M-2n_c-f^{(M)}(\mu))$ all lie within the open interval $(-t,t)$ and so, using the
properties of open intervals, we can again find a small enough $\omega>0$ such that 
$\mu,\mu'_1,\mu'_2\in(-t+\omega,t-\omega)$ and such that 
$\mu + 2E_c (M-2n_c-f^{(M)}(\mu))\in(-t+\omega,t-\omega)$ (i.e., Eq.~\eqref{eq:condition-omega} holds with 
this $\omega$). But this is a contradiction since in this situation our previous results
guarantee the existence of a \emph{unique} $\mu'\in(-t+\omega,t-\omega)$ satisfying
Eq.~\eqref{eq:mu-prime-eqn}. Thus, we have proven the stronger result with $\omega=0$.

To complete the proof we note that, by rearranging Eq.~\eqref{eq:condition-no-omega}, we recover exactly
the condition stated in Eq.~21 of Lemma 1 in the main text.

\subsection{Proof of Theorem 3}

In this section we present the proof of Theorem 3. This proof is very similar to the 
proof of Theorem 1, and so we can actually start from the middle of the proof of Theorem 1. Specifically,
we start from the inequality
\beq
	\delta^{(M)} \geq \delta^{(M)}_0 - \lan \chi^{(M)}_0|\hat{V}|\chi^{(M)}_0\ran\ ,
\eeq
where $\delta^{(M)}= E^{(M)}_1 - E^{(M)}_0$ is the energy gap of $\hat{H}$ in the $M$-particle 
sector, and $\delta^{(M)}_0= \ep^{(M)}_1 - \ep^{(M)}_0$ is the energy gap of $\hat{H}_0$ in the 
$M$-particle sector. For the model at general parameter values, it is difficult to calculate the
expectation value $\lan \chi^{(M)}_0|\hat{V}|\chi^{(M)}_0\ran$ exactly. A useful alternative is
to work with an upper bound for this quantity. Indeed, suppose that we have an upper bound on 
$\lan \chi^{(M)}_0|\hat{V}|\chi^{(M)}_0\ran$, 
\beq
	\lan \chi^{(M)}_0|\hat{V}|\chi^{(M)}_0\ran \leq V_U(M)\ ,
\eeq
where $U$ stands for ``upper.'' Then we find that
\beq
	\delta^{(M)} \geq \delta^{(M)}_0 - V_U(M)\ .
\eeq
This inequality implies that $\hat{H}$ has a unique ground state in the
$M$-particle sector if 
\beq
	\delta^{(M)}_0 > V_U(M)\ .
\eeq

To complete the proof we need to establish the desired upper bound $V_U(M)$ on the expectation value 
$\lan \chi^{(M)}_0|\hat{V}|\chi^{(M)}_0\ran$. For the system with general values of the parameters,
we again indicate the dependence of all quantities on the chemical potential $\mu$. For example,
the total Hamiltonian is $\hat{H}_{\mu}= \hat{H}_{0,\mu}+\hat{V}$, and $|\chi^{(M)}_{0,\mu}\ran$ is the 
ground state of $\hat{H}_{0,\mu}$ in the $M$-particle sector. We also assume that 
$|\mu|<t$ and that $\mu$ and $n_c$ are such 
that the system is at the fine-tuned point
\beq
	M-2n_c= f^{(M)}(\mu) \ .
\eeq
[Recall that $f^{(M)}(\mu)=  \lan \chi^{(M)}_{0,\mu}|\hat{N}_w|\chi^{(M)}_{0,\mu}\ran$.]
We can specialize to this case because, as we know from Lemma 1, there is a large region of the parameter 
space where we can rewrite a Hamiltonian away from this fine-tuned point as a Hamiltonian at this fine-
tuned point (with a different chemical potential), up to an 
overall constant. Then the expectation value that we need to bound is
\beq
	\lan \chi^{(M)}_{0,\mu}|\hat{V}|\chi^{(M)}_{0,\mu}\ran= E_c \lan \chi^{(M)}_{0,\mu}|\left(\hat{N}_w - f^{(M)}(\mu)\right)^2|\chi^{(M)}_{0,\mu}\ran\ .
\eeq
Using Eqs.~\eqref{eq:thm1-bound-even} and 
\eqref{eq:thm1-bound-odd}, we find that
\beq
	\lan \chi^{(M)}_{0,\mu}|\hat{V}|\chi^{(M)}_{0,\mu}\ran \leq \frac{\mathcal{E}_c}{2}\ ,\ M=\text{even}\ ,
\eeq
and
\beq
	\lan \chi^{(M)}_{0,\mu}|\hat{V}|\chi^{(M)}_{0,\mu}\ran \leq \mathcal{E}_c\left(\frac{1}{2}+\frac{8}{L}\right)\ ,\ M=\text{odd}\ .
\eeq
We see that one choice of upper bound that works for all $M$ is
\beq
	V_U(M)= \mathcal{E}_c\left(\frac{1}{2}+\frac{8}{L}\right)\ ,
\eeq
and making this choice completes the proof of Theorem 3. 
We can also see from these results that for the case of even $M$ the result of Theorem 3 holds (for any system size
$L$) under the stronger condition $\mathcal{E}_c < 2\delta^{(M)}_{0,\mu'}$, instead of the slightly weaker
condition $\mathcal{E}_c \left(\frac{1}{2}+\frac{8}{L}\right) < \delta^{(M)}_{0,\mu'}$
that appears in the statement of Theorem 3.

\subsection{Proof of Theorem 4}

In this section we present the proof of Theorem 4. This proof is very similar to the proof of Theorem 2, 
and again relies on a result (Claim 2 below) that
bounds the overlap of the ground state $|\psi^{(M)}_0\ran$ of $\hat{H}$ and the ground state 
$|\chi^{(M)}_0\ran$ of $\hat{H}_0$. 
As in the proof of Theorem 3, the main difficulty that we encounter here is that it is difficult to 
calculate $\lan \chi^{(M)}_0 |\hat{V}|\chi^{(M)}_0\ran$ exactly for the model at general parameter values. 
Therefore, we again assume that we have an upper bound $V_U(M)$ for 
$\lan \chi^{(M)}_0 |\hat{V}|\chi^{(M)}_0\ran$, and we adapt our proof to make use of this upper bound 
instead of the exact value of $\lan \chi^{(M)}_0 |\hat{V}|\chi^{(M)}_0\ran$.

\begin{claim}
Assume that $\delta^{(M)}$ (the gap of $\hat{H}$ in the $M$-particle sector), is positive, 
$\delta^{(M)}>0$. Then 
\begin{align}
	|\lan \chi^{(M)}_0| \psi^{(M)}_0\ran|^2 \geq 1 - \frac{V_U(M)}{\delta^{(M)}_0}\ ,
\end{align}
and this result holds for general values of the parameters $t,\Delta,\mu,\mathcal{E}_c,$ and $n_c$.
\end{claim}

\textbf{Proof of Claim 2:} The result of Claim 2 follows immediately from the result of Claim 1 and the fact 
that $\lan \chi^{(M)}_0 |\hat{V}|\chi^{(M)}_0\ran\leq V_U(M)$.

\textbf{Proof of Theorem 4:} We now use Claim 2 to prove Theorem 4.
As in our proof of Theorem 3, we use a notation that indicates the dependence of the Hamiltonian on the
chemical potential $\mu$, $\hat{H}_{\mu}= \hat{H}_{0,\mu}+\hat{V}$, and we assume that $|\mu|<t$ and that
$\mu$ and $n_c$ satisfy the fine-tuned condition $M-2n_c= f^{(M)}(\mu)$ (recall that restriction
to this case is justified by Lemma 1). Let $\hat{\Gamma}_{\ell,\mu}$ and
$\hat{\Gamma}_{r,\mu}$ be the Majorana-like operators for the Hamiltonian $\hat{H}_{0,\mu}$. 
We know that the operator $i\hat{\Gamma}_{\ell,\mu}^{\dg}\hat{\Gamma}_{r,\mu}$ 
commutes with the projector $\hat{P}_{\chi}= |\chi^{(M)}_{0,\mu}\ran \lan \chi^{(M)}_{0,\mu}|$ onto the 
ground state of $\hat{H}_{0,\mu}$ in the $M$-particle sector.

The proof of Theorem 4 then proceeds in exactly the same manner as the proof of Theorem 2. The only
difference is that the operator $i\hat{A}_1^{\dg}\hat{B}_L$ is replaced by the operator 
$i\hat{\Gamma}_{\ell,\mu}^{\dg}\hat{\Gamma}_{r,\mu}$, and we use the fact that 
$i\hat{\Gamma}_{\ell,\mu}^{\dg}\hat{\Gamma}_{r,\mu}|\chi^{(M)}_{0,\mu}\ran = (-1)^M|\chi^{(M)}_{0,\mu}\ran$. The bound that we find for the Majorana-like correlation function then takes the form
\beq
		(-1)^M\lan \psi^{(M)}_{0,\mu}|i\hat{\Gamma}_{\ell,\mu}^{\dg}\hat{\Gamma}_{r,\mu}|\psi^{(M)}_{0,\mu}\ran \geq 1-2\frac{ V_U(M)}{\delta^{(M)}_{0,\mu}}\  .
\eeq
Finally, just as in the proof of Theorem 3, we use the upper bound
$V_U(M)= \mathcal{E}_c\left(\frac{1}{2}+\frac{8}{L}\right)$, and the result of Theorem 4 then follows 
immediately.

\section{Fermion parity switching}

In this section we prove the existence of a ``fermion parity switch'' in the number-conserving model
in a closed ring geometry. Here, fermion parity switching refers to the phenomenon observed in mean-field
models of TSCs in which the ground state with periodic boundary conditions has odd fermion parity, but
the ground state with anti-periodic boundary conditions has even fermion parity. 
In addition, a
very important property of the mean-field models is that, for both boundary conditions, there
is a finite energy gap between the ground state and the first excited state with the opposite
fermion parity.
We first review this parity switching
phenomenon in the mean-field Kitaev p-wave wire model, 
and then we state and prove a theorem (Theorem 5) that
shows that a fermion parity switch also occurs in our number-conserving model for small enough
$\mathcal{E}_c$. 

We start by reviewing the properties of the mean-field Kitaev p-wave wire Hamiltonian on a closed
ring and with periodic or anti-periodic boundary conditions. The Hamiltonian for this model on a closed
ring takes the form 
\begin{align}
	\hat{H}_{\text{MF}} = -\frac{t}{2}\sum_{j=1}^{L}(\ch^{\dg}_j\ch_{j+1}+\text{h.c.}) - \frac{\Delta}{2}\sum_{j=1}^{L}(\ch_j\ch_{j+1}+\text{h.c.}) 	- \mu  \left(\hat{N}_w -\frac{L}{2}\right)\ ,
\end{align}
where we identify sites $1$ and $L+1$. The first step to diagonalize this Hamiltonian is to go to 
Fourier space by defining
\beq
	\ch_j= \frac{1}{\sqrt{L}}\sum_{k} \ch_k e^{ikj}\ , 
\eeq
where $\ch_k$ are Fourier-transformed fermionic annihilation operators. For periodic boundary conditions the
allowed values of $k$ are contained in the set $\{-\pi,0\}\cup\mathcal{K}_{+}$, where
\beq
	\mathcal{K}_{+}= \left\{\pm\frac{2\pi}{L},\pm\frac{4\pi}{L},\dots,\pm \left(\pi -\frac{2\pi}{L}\right)\right\}\ , \nnb
\eeq
while for anti-periodic boundary conditions the allowed values of $k$ are contained in the set
\beq
\mathcal{K}_{-}=\left\{\pm\frac{\pi}{L},\pm\frac{3\pi}{L},\dots,\pm \left(\pi -\frac{\pi}{L}\right)\right\}\ . \nnb
\eeq
Note that in both cases there are $L$ allowed values of $k$, as there should be. The Fourier-transformed
Hamiltonian takes the form
\beq
	\hat{H}_{\text{MF}}= \sum_k \ep_k \ch^{\dg}_k\ch_k -i\Delta \sum_{k>0} \sin(k)\left( \ch_{-k}\ch_k - \ch^{\dg}_k \ch^{\dg}_{-k} \right) + \mu\frac{L}{2}\ ,
\eeq
where we defined $\ep_k= -t\cos(k)-\mu$. 
In Fourier space we then diagonalize $\hat{H}_{\text{MF}}$ by constructing, for all
$k\neq -\pi,0$, \emph{lowering operators} $\hat{d}_k$ of the form
\beq
	\hat{d}_k= u_k \ch_k + v_k \ch^{\dg}_{-k}\ ,
\eeq
for some coefficients $u_k$ and $v_k$ (these coefficients must also satisfy additional conditions that
guarantee that the $\hat{d}_k$ obey canonical anticommutation relations). 
Demanding that $[\hat{H}_{\text{MF}},\hat{d}_k]= -E_k \hat{d}_k$
for some $E_k\geq0$, we find that $(u_k,v_k)^T$ is the eigenvector of the matrix
$\ep_k\sigma^z + \Delta \sin(k) \sigma^y$ with \emph{positive} eigenvalue, and that
\beq
	E_k= \sqrt{\ep_k^2+\Delta^2\sin^2(k)}\ .
\eeq
For periodic boundary conditions the Hamiltonian can be written (up to an overall constant) as
\beq
	\hat{H}_{\text{MF},+}= \ep_0 \ch^{\dg}_0\ch_0 + \ep_{-\pi}\ch^{\dg}_{-\pi}\ch_{-\pi} + \sum_{k\in\mathcal{K}_{+}}E_k \hat{d}^{\dg}_k\hat{d}_k\ ,
\eeq
while for anti-periodic boundary conditions the Hamiltonian can be written (again up to a constant) as
\beq
	\hat{H}_{\text{MF},-}= \sum_{k\in\mathcal{K}_{-}}E_k \hat{d}^{\dg}_k\hat{d}_k\ .
\eeq
In each case we define the BCS completely paired ground state $|\text{BCS},\pm\ran$ by the condition
$\hat{d}_k|\text{BCS},\pm\ran=0$ for all $k\in\mathcal{K}_{\pm}$, and in the periodic case
we also require $\ch_0|\text{BCS},+\ran=\ch_{-\pi}|\text{BCS},+\ran=0$. This state always has
even fermion parity and takes the familiar BCS form
\beq
	|\text{BCS},\pm\ran= \prod_{k\in\mathcal{K}_{\pm},\ k>0}\left[u_k + v_k \ch^{\dg}_{-k}\ch^{\dg}_k\right]|0\ran\ .
\eeq

We now review the occurrence of a fermion parity switch in this model. To see the switch we first need
to determine the lowest energy states for
periodic and anti-periodic boundary conditions. We start with the anti-periodic case because it is
easier to analyze. In this case, since all $E_k >0$ and there are no unpaired modes to worry about, it is
easy to see that the ground state is $|\text{BCS},-\ran$. In addition, 
the lowest energy excited state is $\hat{d}^{\dg}_{k^*_{-}}|\text{BCS},-\ran$, where 
$k^*_{-}$ is the (possibly non-unique) value of $k$ in $\mathcal{K}_{-}$ that yields the 
lowest possible $E_k$. 
We also define $E_{\text{min},-}= \min_{k\in\mathcal{K}_{-}}E_k$
so that $E_{k^*_{-}}= E_{\text{min},-}$. Thus, we find that for anti-periodic boundary conditions
the ground state has even fermion parity and the energy gap to the first odd parity eigenstate is equal
to $E_{\text{min},-}$.

Next, we consider the case of periodic boundary conditions. This case is complicated by the 
presence of the unpaired fermions at $k=-\pi,0$. Note that $\ep_0= -t-\mu$ and $\ep_{-\pi}=t-\mu$, and
that within the topological phase ($|\mu|<t$ and $\Delta\neq 0$), we have $\ep_0 <0$ and $\ep_{-\pi}>0$. 
As a result, we find that within the topological phase the ground state is $\ch^{\dg}_0|\text{BCS},+\ran$, 
and this state has odd fermion parity. Next, depending on the
specific parameter values, the first excited state (which has even fermion parity) can be any one
of the states $|\text{BCS},+\ran$, $\ch^{\dg}_0\ch^{\dg}_{-\pi}|\text{BCS},+\ran$, or
$\ch^{\dg}_0\hat{d}^{\dg}_{k^*_{+}}|\text{BCS},+\ran$, where $k^*_{+}$
is the (possibly non-unique) value of $k$ in $\mathcal{K}_{+}$ that yields the lowest possible $E_k$. 
We also define $E_{\text{min},+}= \min_{k\in\mathcal{K}_{+}}E_k$, 
so that $E_{k^*_{+}}= E_{\text{min},+}$. The energy gap between the ground
state $\ch^{\dg}_0|\text{BCS},+\ran$ and the first even parity excited state in these three cases is 
equal to $t+\mu$, $t-\mu$, and $E_{\text{min},+}$, respectively. Thus, we find that for periodic boundary 
conditions the ground state has odd fermion parity and the energy gap to the first even parity
eigenstate is equal to $\min\left\{t+\mu,t-\mu,E_{\text{min},+}\right\}$.

In both cases (periodic and anti-periodic boundary conditions) the gap between the ground state and
the first excited state is bounded from below by $E_{\text{min}}=\min_{k\in[-\pi,\pi)}E_k$, where
the minimum here is taken over the continuum of $k$ values in the interval $[-\pi,\pi)$. This is 
clear in the case with anti-periodic boundary conditions, as $\mathcal{K}_{-}\subset [-\pi,\pi)$, 
and so $E_{\text{min},-}\geq E_{\text{min}}$. To see this in the case of periodic boundary
conditions, we need to show that $t+\mu\geq E_{\text{min}}$ and that $t-\mu\geq E_{\text{min}}$ (it
is clear that $E_{\text{min,+}}\geq E_{\text{min}}$ since $\mathcal{K}_{+}\subset [-\pi,\pi)$).
We can prove this by noting that $dE_k/dk=0$ at $k=0,-\pi$, and so $k=0,-\pi$ are the locations
of minima, maxima, or inflection points of $E_k$. 
No matter which possibility occurs, the minimum value of $E_k$ must 
always be less than or equal to $E_0$ and $E_{-\pi}$. Then, since $E_0= t+\mu$ and $E_{-\pi}=t-\mu$ 
(for $|\mu|<t$), we find that $t-\mu\geq E_{\text{min}}$ and that 
$t+\mu\geq E_{\text{min}}$.

To summarize, for the mean-field Hamiltonian $\hat{H}_{\text{MF}}$ within the topological phase
($|\mu|<t$ and $\Delta\neq 0$), the lowest energy state with periodic 
boundary conditions has odd fermion parity, while the lowest
energy state with anti-periodic boundary conditions has even fermion parity.
In addition, for both boundary conditions the gap between the ground 
state and the first excited state with the opposite fermion parity is bounded from 
below by $E_{\text{min}}=\min_{k\in[-\pi,\pi)}E_k$. 

We now move on to the number-conserving case and introduce some notation that will allow us to 
state our result on the fermion parity switch in the number-conserving model. We again
use a notation that indicates the dependence of all quantities on $\mu$. Recall that in the 
$M$-particle sector the charging energy term can be rewritten as
\beq
	\hat{V}= E_c\left[\hat{N}_w - (M - 2n_c)\right]^2\ .
\eeq
We can rewrite this as
\beq
	\hat{V}= E_c\left[\hat{N}_w - f^{(M)}(\mu)\right]^2 + 2E_c s(M,\mu)\left(\hat{N}_w - f^{(M)}(\mu)\right) + E_c s(M,\mu)^2\ ,
\eeq
where 
\beq
	s(M,\mu)= f^{(M)}(\mu)-(M-2n_c)\ .
\eeq
Note that $s(M,\mu)$ is the difference between $f^{(M)}(\mu)$, the average fermion number on the
wire in the ground state $|\chi^{(M)}_{0,\mu}\ran$ of $\hat{H}_{0,\mu}$, and $M-2n_c$, the filling
preferred by the charging energy term in the $M$-particle sector. For later use, we note that 
(for even $L$) we
can use Eqs.~\eqref{eq:thm1-bound-even} and \eqref{eq:thm1-bound-odd} to obtain the upper bounds
\beq
	\lan\chi^{(M)}_{0,\mu}|\hat{V}|\chi^{(M)}_{0,\mu}\ran \leq \mathcal{E}_c\left(\frac{1}{2}+
	\frac{s(M,\mu)^2}{L}\right)\ ,\ M = \text{even}\ , \label{eq:even-M-s-bound}
\eeq
and
\beq
	\lan\chi^{(M)}_{0,\mu}|\hat{V}|\chi^{(M)}_{0,\mu}\ran \leq \mathcal{E}_c\left(\frac{1}{2}+\frac{8}{L}+\frac{s(M,\mu)^2}{L}\right)\ ,\ M = \text{odd}\ . \label{eq:odd-M-s-bound}
\eeq

Finally, we define 
\begin{subequations}
\beqa
	s_{+}(\mu) &=& \min_{\text{even }M}|s(M,\mu)| \\
	s_{-}(\mu) &=& \min_{\text{odd }M}|s(M,\mu)|\ .
\eeqa 
\end{subequations}
For either sign (``$+$'' or ``$-$'') we have the inequality
\beq
	s_{\pm}(\mu) \leq 1\ ,
\eeq
and this can be seen as follows. First, consider the case of even $M$ and note that we can always write 
$f^{(M)}(\mu)+2n_c= p+\al$ for some even
integer $p$ and some real number $\al\in[-1,1]$ (indeed, \emph{any} real number can
be written in this way). The even integer $p$ appearing here is also independent of
the specific choice of $M$ because $f^{(M)}(\mu)$ itself only depends on the parity of $M$ (i.e., 
$f^{(M_1)}(\mu)=f^{(M_2)}(\mu)$ if $(-1)^{M_1}=(-1)^{M_2}$). By choosing $M=p$ we obtain
the bound $s_{+}(\mu)\leq |s(p,\mu)| = |\al|$, and so $s_{+}(\mu) \leq 1$. A similar argument
but with $p$ odd shows that $s_{-}(\mu) \leq 1$ as well.

With this notation in hand we can now state our theorem on the existence of 
a fermion parity switch in the number-conserving model. For technical reasons we again assume here 
that $L$ is even.

\begin{thm}[Fermion parity switch]
Let $E_k= \sqrt{(t\cos(k)+\mu)^2+\Delta^2\sin^2(k)}$ and let $E_{\text{min}}=\min_{k\in[-\pi,\pi)}E_k$. 
If $\mathcal{E}_c\left(\frac{1}{2}+\frac{9}{L}\right) < E_{\text{min}}$, then the overall ground state
of $\hat{H}_{\mu}$ with periodic (anti-periodic) boundary conditions occurs in a sector with
odd (even) total particle number. In addition, in both cases $\delta_{\text{parity}}$, the energy gap
between the overall ground state of $\hat{H}_{\mu}$ and the lowest energy eigenstate with opposite number 
parity, is bounded from below as
\beq
	\delta_{\text{parity}} \geq E_{\text{min}}-\mathcal{E}_c\left(\frac{1}{2}+\frac{9}{L}\right) =  E_{\text{min}}-\frac{\mathcal{E}_c}{2} + O\left(\frac{1}{L}\right)\ ,
\eeq
and this bound holds for all system sizes $L$.
\end{thm}

\subsection{Proof of Theorem 5}

The proof of Theorem 5 is very similar to the proofs of Theorems 1 and 3 
and again relies on the positivity of $\hat{V}$. To start,
recall that the energy spectrum of $\hat{H}_{0,\mu}$ in the 
$M$-particle sector is identical to the energy spectrum of $\hat{H}_{\text{MF},\mu}$ in the sector with
fermion parity $(-1)^M$. As a result, $\hat{H}_{0,\mu}$ has the property that its 
overall ground state with periodic boundary conditions occurs in a sector with odd total particle 
number, while its overall ground state with anti-periodic boundary conditions occurs in a sector with
even total particle number. 
In fact, this property of the spectrum of $\hat{H}_{0,\mu}$ implies a much stronger result. Let
$M_e$ be any even integer, and let $M_o$ be any odd integer. Then for 
periodic boundary conditions we have $\ep^{(M_e)}_{0,\mu} - \ep^{(M_o)}_{0,\mu} \geq E_{\text{min}}$,
where $\ep^{(M)}_{0,\mu}$ is the ground state energy of $\hat{H}_{0,\mu}$ in the $M$-particle sector.
In the anti-periodic case the situation is reversed and we instead have 
$\ep^{(M_o)}_{0,\mu} - \ep^{(M_e)}_{0,\mu} \geq E_{\text{min}}$. 

Next, we turn to the full Hamiltonian $\hat{H}_{\mu}$ with the charging energy term, and we
consider the case of periodic boundary conditions. The positivity
of $\hat{V}$ implies that $\ep^{(M)}_{0,\mu} \leq E^{(M)}_{0,\mu}$ for any $M$, 
where $E^{(M)}_{0,\mu}$ is the ground state energy of $\hat{H}_{\mu}$ in the $M$-particle sector. 
As in the proofs of Theorems 1 and 3, combining this property with the variational theorem
for the ground state energy in the $M_o$-particle sector yields the inequality
\beq
	E^{(M_e)}_{0,\mu} - E^{(M_o)}_{0,\mu} \geq \ep^{(M_e)}_{0,\mu} - \ep^{(M_o)}_{0,\mu} - \lan\chi^{(M_o)}_{0,\mu}|\hat{V}|\chi^{(M_o)}_{0,\mu}\ran\ .
\eeq
Next, using Eq.~\eqref{eq:odd-M-s-bound} and the fact that 
$\ep^{(M_e)}_{0,\mu} - \ep^{(M_o)}_{0,\mu} \geq E_{\text{min}}$, we obtain
\beq
	E^{(M_e)}_{0,\mu} - E^{(M_o)}_{0,\mu} \geq E_{\text{min}} - \mathcal{E}_c\left(\frac{1}{2}+\frac{8}{L}+\frac{s(M_o,\mu)^2}{L}\right)\ .
\eeq
This bound holds for any $M_o$ and, in particular, holds for the value $M^*_o$ such that 
$|s(M^*_o,\mu)|=s_{-}(\mu)$ (if $M_o^*$ is not unique then we just choose a particular one). 
Therefore we obtain the bound
\beq
	E^{(M_e)}_{0,\mu} - E^{(M^*_o)}_{0,\mu} \geq E_{\text{min}} - \mathcal{E}_c\left(\frac{1}{2}+\frac{8}{L}+\frac{s_{-}(\mu)^2}{L}\right)\ ,
\eeq
and then using $s_{-}(\mu) \leq 1$ gives
\beq
	E^{(M_e)}_{0,\mu} - E^{(M^*_o)}_{0,\mu} \geq E_{\text{min}} - \mathcal{E}_c\left(\frac{1}{2}+\frac{9}{L}\right)\ .
\eeq
We find that, if 
$E_{\text{min}} > \mathcal{E}_c\left(\frac{1}{2}+\frac{9}{L}\right)$, then 
the energy of the ground state of $\hat{H}_{\mu}$ in the $M_o^*$-particle sector is strictly less than the
energy of the ground state in \emph{any} even particle number sector. This proves that, if
$E_{\text{min}} > \mathcal{E}_c\left(\frac{1}{2}+\frac{9}{L}\right)$, then
the overall ground state of $\hat{H}_{\mu}$ with periodic boundary conditions occurs in a sector
with odd total particle number.

In the case of anti-periodic boundary conditions, similar reasoning leads to the inequality
\beq
	E^{(M_o)}_{0,\mu} - E^{(M_e)}_{0,\mu} \geq \ep^{(M_o)}_{0,\mu} - \ep^{(M_e)}_{0,\mu} - \lan\chi^{(M_e)}_{0,\mu}|\hat{V}|\chi^{(M_e)}_{0,\mu}\ran\ ,
\eeq
where the roles of $M_e$ and $M_o$ are now reversed as compared with the periodic case.
We now use Eq.~\eqref{eq:even-M-s-bound} and  
$\ep^{(M_o)}_{0,\mu} - \ep^{(M_e)}_{0,\mu} \geq E_{\text{min}}$ to obtain
\beq
	E^{(M_o)}_{0,\mu} - E^{(M_e)}_{0,\mu} \geq E_{\text{min}} - \mathcal{E}_c\left(\frac{1}{2}+\frac{s(M_e,\mu)^2}{L}\right)\ .
\eeq
This bound holds for any $M_e$ and, in particular, holds for the value $M^*_e$ such that 
$|s(M^*_e,\mu)|=s_{+}(\mu)$ (again, if $M^*_e$ is not unique then we just pick a particular one).
This yields the bound
\beq
	E^{(M_o)}_{0,\mu} - E^{(M^*_e)}_{0,\mu} \geq E_{\text{min}} - \mathcal{E}_c\left(\frac{1}{2}+\frac{s_{+}(\mu)^2}{L}\right)\ ,
\eeq
and then using $s_{+}(\mu) \leq 1$ gives
\beq
	E^{(M_o)}_{0,\mu} - E^{(M^*_e)}_{0,\mu} \geq E_{\text{min}} - \mathcal{E}_c\left(\frac{1}{2}+\frac{1}{L}\right) \geq E_{\text{min}} - \mathcal{E}_c\left(\frac{1}{2}+\frac{9}{L}\right)\ .
\eeq
The final inequality here is not strictly necessary, but we find it useful as it allows us to use the same 
bound as in the case of periodic boundary conditions.
We then find that, if 
$E_{\text{min}} > \mathcal{E}_c\left(\frac{1}{2}+\frac{9}{L}\right)$, then 
the energy of the ground state of $\hat{H}_{\mu}$ in the $M_e^*$-particle sector is strictly less than
the energy of the ground state in \emph{any} odd particle number sector. This proves that, if
$E_{\text{min}} > \mathcal{E}_c\left(\frac{1}{2}+\frac{9}{L}\right)$, then
the overall ground state of $\hat{H}_{\mu}$ with anti-periodic boundary conditions occurs in a sector
with even total particle number.

Finally, in both cases $\delta_{\text{parity}}$,
the energy gap between the overall ground state and the lowest energy eigenstate with opposite number 
parity, is bounded from below as
\beq
	\delta_{\text{parity}} \geq E_{\text{min}}-\mathcal{E}_c\left(\frac{1}{2}+\frac{9}{L}\right) =  E_{\text{min}}-\frac{\mathcal{E}_c}{2} + O\left(\frac{1}{L}\right)\ .
\eeq

\section{Number-conserving model without charging energy}
\label{app:Ec-zero}

In this section we explain the relation between the number-conserving TSC model with
no charging energy ($E_c=0$ so $\hat{V}=0$), and the more familiar mean-field TSC model. 
In particular, we show how the eigenstates of the number-conserving model at fixed total particle number 
$\hat{N}=M$ can be written in terms of the eigenstates of the mean-field model with Hamiltonian 
$\hat{H}_{\text{MF}}$. We also show how certain expectation values in the number-conserving model with $E_c=0$
can be expressed in terms of expectation values in the mean-field model.

To start, let $\hat{H}_{\text{MF}}(\phi)$ denote the mean-field Hamiltonian obtained from $\hat{H}_0$ after replacing 
the quantum operator $\ph$ with the classical phase $\phi$, 
\begin{align}
	\hat{H}_{\text{MF}}(\phi) =& -\frac{t}{2}\sum_{j=1}^{L-1}(\ch^{\dg}_j\ch_{j+1}+\text{h.c.}) - \frac{\Delta}{2}\sum_{j=1}^{L-1}(\ch_j\ch_{j+1}e^{i\phi}+\text{h.c.}) \nnb \\
	-&\ \mu  \left(\hat{N}_w -\frac{L}{2}\right)\ .
\end{align}
Let $|\chi_a(\phi)\ran$ (for some range of the index $a$) be a complete set of eigenstates of this mean-field fermionic 
Hamiltonian, and let $\ep_a$ be the energies of these states, 
$\hat{H}_{\text{MF}}(\phi)|\chi_a(\phi)\ran= \ep_a |\chi_a(\phi)\ran$. 
The energies $\ep_a$ are independent of $\phi$ because $\hat{H}_{\text{MF}}(\phi)$ is related to 
$\hat{H}_{\text{MF}}(0)=\hat{H}_{\text{MF}}$ by a unitary 
transformation. Indeed, we have
\beq
	\hat{H}_{\text{MF}}(\phi)= e^{-i\frac{\phi}{2}\hat{N}_w}\hat{H}_{\text{MF}}(0) e^{i\frac{\phi}{2}\hat{N}_w}\ ,
\eeq
and this implies that
\beq
	|\chi_a(\phi)\ran= e^{-i\frac{\phi}{2}\hat{N}_w}|\chi_a(0)\ran\ . \label{eq:chi-phi}
\eeq

We now show how to construct eigenstates $|\chi^{(M)}_a\ran$ of $\hat{H}_0$ at fixed total particle
number $\hat{N}=M$ in terms of the eigenstates $|\chi_a(\phi)\ran$ of the fermionic mean-field Hamiltonian
$\hat{H}_{\text{MF}}(\phi)$. To start, recall the eigenstates $|p\ran$, $p\in\mathbb{Z}$, of the Cooper
pair number operator $\hat{n}$, which span the Hilbert space $\mathcal{H}_{\text{SC}}$ of the number and phase degree
of freedom. These states obey
\begin{subequations}
\beqa
	\lan p|p'\ran &=& \delta_{p,p'} \\
	\hat{n}|p\ran &=& p |p\ran\ .
\eeqa
\end{subequations}
We also introduce the eigenstates $|\phi\ran$ of the phase operator $\hat{\phi}$. These states obey
\begin{subequations}
\beqa
	\lan\phi|\phi'\ran &=& \delta(\phi-\phi') \\
	\hat{\phi}|\phi\ran &=& \phi |\phi\ran\ .
\eeqa
\end{subequations}
We also have the inner product
\beq
	\lan \phi |p\ran= \frac{1}{\sqrt{2\pi}}e^{i p\phi}\ ,
\eeq
which allows us to write $|\phi\ran$ as
\beq
	|\phi\ran= \sum_{p\in\mathbb{Z}}|p\ran \lan p|\phi\ran = \sum_{p\in\mathbb{Z}}\frac{e^{-i p\phi}}{\sqrt{2\pi}}|p\ran \ . \label{eq:phi-p}
\eeq

Using these basis states, we now show that the eigenstates $|\chi^{(M)}_a\ran$ of $\hat{H}_0$ at fixed total particle
number $M$ take the form
\beq
	|\chi^{(M)}_a\ran= \frac{1}{\sqrt{2\pi}}\int_0^{2\pi}d\phi\ e^{i\frac{M}{2}\phi}|\phi\ran \otimes |\chi_a(\phi)\ran\ . \label{eq:fixed-number-states}
\eeq
These states are normalized if we assume the normalization $\lan\chi_a(\phi)|\chi_b(\phi)\ran = \delta_{ab}$ for the
eigenstates of the mean-field fermionic Hamiltonian. In addition, there is one constraint on these
states, which is that $(-1)^{M}$ must be equal to the fermion parity of the state $|\chi_a(0)\ran$, otherwise
the integral over $\phi$ which yields the state $|\chi^{(M)}_a\ran$ will actually evaluate to zero. 
This constraint makes sense since the
Cooper pairs have even fermion parity, so $(-1)^{\hat{N}}=(-1)^{\hat{N}_w}$.

We now show that $|\chi^{(M)}_a\ran$ is an eigenstate of $\hat{H}_0$ with the same eigenvalue
$\ep_a$ as the eigenstate $|\chi_a(\phi)\ran$ of $\hat{H}_{\text{MF}}(\phi)$.
To prove that $\hat{H}_0|\chi^{(M)}_a\ran= \ep_a |\chi^{(M)}_a\ran$, we simply note that 
\beqa
	\hat{H}_0|\phi\ran \otimes |\chi_a(\phi)\ran &=& \hat{H}_{\text{MF}}(\phi)|\phi\ran \otimes |\chi_a(\phi)\ran \nnb \\
	&=& \ep_a |\phi\ran \otimes |\chi_a(\phi)\ran \ ,
\eeqa
where we used the fact that $\hat{\phi}|\phi\ran = \phi |\phi\ran$ and then the fact that 
$\hat{H}_{\text{MF}}(\phi)|\chi_a(\phi)\ran=\ep_a|\chi_a(\phi)\ran$.

The next property to prove is that $\hat{N}|\chi^{(M)}_a\ran= M|\chi^{(M)}_a\ran$, i.e., that $|\chi^{(M)}_a\ran$ 
is an eigenstate of the total particle number operator $\hat{N}=\hat{N}_w+2\hat{n}$ with eigenvalue $M$. To prove this we 
first use Eq.~\eqref{eq:phi-p} and Eq.~\eqref{eq:chi-phi} to rewrite $|\chi^{(M)}_a\ran$ in the form
\beq
	|\chi^{(M)}_a\ran= \frac{1}{2\pi}\sum_{p\in\mathbb{Z}}\int_0^{2\pi}d\phi\ e^{i\left(\frac{M}{2}-\frac{\hat{N}_w}{2}\right)\phi}e^{-ip\phi}|p\ran \otimes |\chi_a(0)\ran\ .
\eeq 
Next, we apply the operator $\hat{n}$ to find 
\beqa
	\hat{n}|\chi^{(M)}_a\ran &=& \frac{1}{2\pi}\sum_{p\in\mathbb{Z}}\int_0^{2\pi}d\phi\ e^{i\left(\frac{M}{2}-\frac{\hat{N}_w}{2}\right)\phi}e^{-ip\phi} p |p\ran \otimes |\chi_a(0)\ran \nnb \\
	&=& i \frac{1}{2\pi}\sum_{p\in\mathbb{Z}}\int_0^{2\pi}d\phi\ e^{i\left(\frac{M}{2}-\frac{\hat{N}_w}{2}\right)\phi} \frac{\pd}{\pd\phi} \left[e^{-ip\phi}\right]  |p\ran \otimes |\chi_a(0)\ran\nnb \\
	&=& -i \frac{1}{2\pi}\sum_{p\in\mathbb{Z}}\int_0^{2\pi}d\phi\ \frac{\pd}{\pd\phi}\left[e^{i\left(\frac{M}{2}-\frac{\hat{N}_w}{2}\right)\phi} \right] e^{-ip\phi}  |p\ran \otimes |\chi_a(0)\ran \nnb \\
	&=& \left(\frac{M}{2}-\frac{\hat{N}_w}{2}\right)|\chi^{(M)}_a\ran\ ,
\eeqa
which is equivalent to the desired result that $(\hat{N}_w+2\hat{n})|\chi^{(M)}_a\ran=M|\chi^{(M)}_a\ran$.
Note that in this derivation we needed to integrate by parts to get from the second line to the third line. 
In this integration by parts the boundary term at $\phi=2\pi$ cancels the boundary term at $\phi=0$ because of the 
constraint that $(-1)^{M}$ is equal to the fermion parity of the mean-field eigenstate $|\chi_a(0)\ran$, i.e., we 
needed to use the fact that $e^{i2\pi\left(\frac{M}{2}-\frac{\hat{N}_w}{2}\right)} |\chi_a(0)\ran=  |\chi_a(0)\ran$.

We close this section by discussing relations between expectation values of operators in the
eigenstates $|\chi^{(M)}_a\ran$ and the mean-field eigenstates $|\chi_a(0)\ran$. Consider any operator
$\hat{\mathcal{O}}$ that can be written in terms of the fermionic creation and annihilation operators
$\ch_j$ and $\ch_j^{\dg}$ and that also commutes with fermion number operator $\hat{N}_w$, 
$[\hat{\mathcal{O}},\hat{N}_w]=0$. Then one can easily show (using the explicit expression 
\eqref{eq:fixed-number-states} for $|\chi^{(M)}_a\ran$) that
\beq
	\lan \chi^{(M)}_a| \hat{\mathcal{O}} |\chi^{(M)}_a\ran = \lan \chi_a(0)| \hat{\mathcal{O}} |\chi_a(0)\ran\ ,
\eeq
i.e., the expectation value of $\hat{\mathcal{O}}$ in the eigenstate $|\chi^{(M)}_a\ran$ of
$\hat{H}_0$ is equal to the expectation value of $\hat{\mathcal{O}}$ in the mean-field state $|\chi_a(0)\ran$.
We can also derive similar relations for operators that do not commute with $\hat{N}_w$. For example, one
can show that
\beq
	\lan \chi^{(M)}_a| \ch_j \ch_k e^{i\ph} |\chi^{(M)}_a\ran = \lan \chi_a(0)|  \ch_j \ch_k |\chi_a(0)\ran\ .
\eeq
The insertion of the phase operator $e^{i\ph}$ on the left-hand side of this equation allows the expectation
value to be non-zero even for the state $|\chi^{(M)}_a\ran$ (which has a fixed total particle number equal to 
$M$).

\section{Properties of the mean-field model at $t=\Delta$ and $\mu=0$}
\label{app:JW}

In this section we review some basic properties of the mean-field TSC model $\hat{H}_{\text{MF}}$ at the
special point $t=\Delta$ and $\mu=0$. Recall that we used the notation 
$\hat{H}_{\text{MF},*}$ for the mean-field Hamiltonian at this special point. 
The properties of this model are well-known in the literature but we find it useful to collect them all in 
one place here. Our main tool to derive these properties is the
Jordan-Wigner (JW) transformation, which can be used to convert the fermionic wire model to a spin model. 
The JW transformation for the Majorana fermion operators is as follows. We have
\begin{subequations}
\beqa
	\hat{a}_j &=& \left(\prod_{k<j}\hat{\sigma}^x_k \right)\hat{\sigma}^z_j \\
	\hat{b}_j &=& \left(\prod_{k<j}\hat{\sigma}^x_k \right)\hat{\sigma}^y_j \ .
\eeqa
\end{subequations}
where the operator $\hat{\sigma}^a_j$ (for $a\in\{x,y,z\}$) acts as the Pauli matrix $\sigma^a$ on site $j$ 
and as the identity on all other sites. Using this transformation we find that $i\hat{a}_j\hat{b}_j= \hat{\sigma}^x_j$ 
and that $i\hat{b}_j\hat{a}_{j+1}= \hat{\sigma}^z_j \hat{\sigma}^z_{j+1}$. Using these results we 
find that after the JW transformation $\hat{H}_{\text{MF},*}$ takes the form
\beq
	\hat{H}_{\text{MF},*}= - \frac{\Delta}{2}\sum_{j=1}^{L-1}\hat{\sigma}^z_j \hat{\sigma}^z_{j+1}\ ,
\eeq
i.e., the JW-transformed Hamiltonian is that of a \emph{classical} Ising model with open boundaries. It is easy to see that
this Hamiltonian has two degenerate ground states, namely the states $|\Uparrow\ran$ and $|\Downarrow\ran$,
where $|\Uparrow\ran$ is the state of the spin chain with all spins pointing up 
($\hat{\sigma}^z_j |\Uparrow\ran= |\Uparrow\ran\ \forall\ j$) and $|\Downarrow\ran$ is the state of the spin chain with all 
spins pointing down ($\hat{\sigma}^z_j |\Downarrow\ran= -|\Downarrow\ran\ \forall\ j$).
We also find that the number operator for the fermions on the wire takes the form
\beq
	\hat{N}_w= \frac{L}{2} + \frac{1}{2}\sum_{j=1}^L \hat{\sigma}^x_j\ .
\eeq

For the system in the spin language we can define an Ising symmetry operator $\hat{S}$ by
\beq
	\hat{S}= \prod_{j=1}^L \hat{\sigma}^x_j\ ,
\eeq
and we can see that $\hat{S}^2=1$ and that $[\hat{S},\hat{H}_{\text{MF},*}]=0$. In fact, $\hat{S}$ is, up 
to a possible minus sign, the JW-transform of the fermion parity operator,
\beq
	\hat{S}= (-1)^L (-1)^{\hat{N}_w}\ .
\eeq
From now on we assume that $L$ is even, and in this case we have
$\hat{S}= (-1)^{\hat{N}_w}$.
Since physical states have definite fermion parity, when studying $\hat{H}_{\text{MF},*}$ in the spin language we should 
choose the eigenvectors to also be eigenvectors of $\hat{S}$. This means that the two degenerate ground states
$|\pm\ran$ of $\hat{H}_{\text{MF},*}$ correspond, in the spin language, to the two states
\beq
	|\pm\ran = \frac{1}{\sqrt{2}}\left( |\Uparrow\ran \pm |\Downarrow\ran\right)\ .
\eeq

We now briefly discuss the excited states of the model. It is clear that excited states in the even or odd 
parity sectors can be constructed by acting on the ground states $|\pm\ran$ with $\hat{\sigma}^x_j$ 
operators. In this way one can easily see that the energy gap of $\hat{H}_{\text{MF},*}$ in both the even and 
odd parity sectors is equal to $\Delta$. The states $\hat{\sigma}^x_1 |\pm\ran$ are examples of excited 
states in the even and odd parity sectors with the lowest possible energy. 

Next, we study the generating functions for expectation values of powers of the operator
$\hat{N}_w - \frac{L}{2}$ in the two ground states $|\pm\ran$. Note that the operator 
$\hat{N}_w - \frac{L}{2}$ itself has zero expectation value in the states $|\pm\ran$. 
We define the generating functions
\beq
	f_{\pm}(\tau)= \lan \pm | e^{\tau (\hat{N}_w-\frac{L}{2})} |\pm\ran\ .
\eeq
Using these generating functions, we can compute expectation values of powers of $\hat{N}_w - \frac{L}{2}$ by
taking derivatives with respect to $\tau$, 
\beq
	\lan\pm| \left(\hat{N}_w - \frac{L}{2}\right)^p |\pm\ran = \frac{d^p f_{\pm}(\tau)}{d\tau^p}\Bigg|_{\tau=0}\ .
\eeq
Using the JW-transform we find that
\beqa
	f_{\pm}(\tau)&=& \lan\pm| e^{\frac{\tau}{2}\sum_{j=1}^L\hat{\sigma}^x_j} |\pm\ran \nnb \\
		&=& \lan\pm| \prod_{j=1}^L\left[ \cosh\left(\frac{\tau}{2}\right) + \sinh\left(\frac{\tau}{2}\right)\hat{\sigma}^x_j\right]|\pm\ran \nnb \\
		&=& \lan\pm| \left[ \cosh\left(\frac{\tau}{2}\right)^L + \sinh\left(\frac{\tau}{2}\right)^L \hat{S}\right]|\pm\ran \nnb \\
		&=& \cosh\left(\frac{\tau}{2}\right)^L \pm \sinh\left(\frac{\tau}{2}\right)^L\ . \label{eq:gen-func}
\eeqa
Here, to get from the second to the third line, we used the fact that the expectation value of a product of 
$\hat{\sigma}^x_j$'s in one of the states $|\pm\ran$ can only be non-zero if there is a $\hat{\sigma}^x_j$ for 
every single site in the wire. We also used $\hat{S}|\pm\ran= \pm|\pm\ran$ to get from the third line to the fourth line.

For use in the main text, we record here some expectation values of powers of $\hat{N}_w - \frac{L}{2}$
that can be derived using these generating functions. Let us assume that $L > 2$. Then we have
\begin{subequations}
\label{eq:spin-chain-expectations}
\beqa
	\lan \pm |\left(\hat{N}_w - \frac{L}{2} \right)|\pm\ran &=& 0 \\
	\lan \pm |\left(\hat{N}_w - \frac{L}{2} \right)^2|\pm\ran &=& \frac{L}{4} \ .
\eeqa
\end{subequations}
In general, one finds that $\lan +|\left(\hat{N}_w - \frac{L}{2} \right)^p|+\ran =
\lan - |\left(\hat{N}_w - \frac{L}{2} \right)^p|-\ran$ for $p<L$. This can be seen from the fact that the difference
of the two generating functions takes the form $f_{+}(\tau)-f_{-}(\tau)= 2\sinh\left(\frac{\tau}{2}\right)^L$, and the
first term in the Taylor expansion of this difference about $\tau=0$ is of order $\tau^L$.

We close this section by explaining the connection between the fermion parity of the ground states 
$|\pm\ran$ of $\hat{H}_{\text{MF},*}$ and the eigenvalues of the operator $i\hat{a}_1\hat{b}_L$ when 
acting on these states. We again assume that $L$ is even. Using the JW transformation we find that 
\beqa
	i\hat{a}_1\hat{b}_L &=& i\hat{\sigma}^z_1 \left(\prod_{j=1}^{L-1}\hat{\sigma}^x_j\right)\hat{\sigma}^y_L \nnb \\
	&=& \hat{\sigma}^z_1\hat{\sigma}^z_L \hat{S}\ .
\eeqa
Then, since $\hat{\sigma}^z_1\hat{\sigma}^z_L|\pm\ran = |\pm\ran$, we find that
\beq
	i\hat{a}_1\hat{b}_L |\pm\ran = \pm|\pm\ran\ ,
\eeq
i.e., the operator $i\hat{a}_1\hat{b}_L$ is equal to the fermion parity operator $(-1)^{\hat{N}_w}$ within
the ground state subspace.

\section{Particle number expectation values in mean-field models of superconductivity}
\label{app:number-bounds}

In this section we derive rigorous upper bounds on expectation values of powers of the particle number
operator $\hat{N}_w=\sum_{j=1}^L \ch^{\dg}_j\ch_j$ in ground states of mean-field models of 
superconductivity. These bounds are an essential ingredient in our proofs of Theorems 3, 4, and 5.
The formalism needed for dealing with general mean-field models of superconductivity (e.g.,
models without translation-invariance) can be found in Sec.~II of Ref.~\onlinecite{read2009} and
in Appendix A of Ref.~\onlinecite{stone-chung} (note, however, that our conventions do not match exactly
with the conventions in these references). We also require certain general results about complex 
antisymmetric matrices. A useful reference for this material in a physics context is Appendix A of 
Ref.~\onlinecite{yang-ODLRO} (note that in this reference a tilde is used to denote the transpose of a matrix 
or a vector).

Hamiltonians for general mean-field models of superconductivity have the form
\beq
	\hat{H}_{\text{MF}}= \sum_{i,j=1}^L\left( h_{ij}\ch^{\dg}_i\ch_j + \frac{1}{2}\Delta_{ij}\ch^{\dg}_i\ch^{\dg}_j +\frac{1}{2}\ov{\Delta}_{ij}\ch_j\ch_i\right)\ ,
\eeq
where $h_{ij}$ and $\Delta_{ij}$ are the matrix elements of two $L\times L$ matrices $h$ and $\Delta$, and
these matrices completely define the model (we also need $h_{ji}=\ov{h}_{ij}$ so that the Hamiltonian is
Hermitian). In what follows we also assume that $L$ is even.

The standard procedure to diagonalize Hamiltonians of this form is to construct lowering operators 
$\hat{d}_i$ that obey $[\hat{H}_{\text{MF}},\hat{d}_i]=-E_i \hat{d}_i$, for some energy eigenvalues
$E_i \geq 0$. The Hamiltonian can then be rewritten as
\beq
	\hat{H}_{\text{MF}}= \sum_{i=1}^L E_i \hat{d}^{\dg}_i\hat{d}_i + \text{constant}\ ,
\eeq
and the exact form of the constant term will not be important for us here. 
The $\hat{d}_i$ operators take the general form 
\beq
	\hat{d}_i= \sum_{j=1}^L (u_{ij}\ch_j + v_{ij}\ch^{\dg}_j)\ ,
\eeq
where $u_{ij}$ and $v_{ij}$ are the elements of two new $L\times L$ matrices $u$ and $v$. 
These matrices are determined by solving the equations $[\hat{H}_{\text{MF}},\hat{d}_i]=-E_i \hat{d}_i$ for 
all $i$. These equations are a generalized version of the Bogoliubov-de Gennes equations, and one can 
think of the operator $\hat{d}_i$ as a generalized version of a Bogoliubov quasiparticle operator. One
can also derive additional relations that $u$ and $v$ must satisfy so that the $\hat{d}_i$ operators obey standard
anticommutation relations. In what follows, we assume that these additional relations are satisfied
so that $\{\hat{d}_i,\hat{d}_j\}=0$ and $\{\hat{d}_i,\hat{d}^{\dg}_j\}=\delta_{ij}$. 

The even parity ground state $|+\ran$ of $\hat{H}_{\text{MF}}$ is defined to satisfy
$\hat{d}_i|+\ran = 0\ \forall\ i$. One can show that this state takes the form
\beq
	|+\ran= \mathcal{N}e^{\hat{G}}|0\ran\ ,
\eeq
where
\beq
	\hat{G}=\frac{1}{2}\sum_{i,j}g_{ij}\ch^{\dg}_i\ch^{\dg}_j\ ,
\eeq
and $g_{ij}$ are the elements of a complex antisymmetric matrix $g$ related to the 
matrices $u$ and $v$ by
\beq
	g= -u^{-1}v\ .
\eeq
In addition, $|0\ran$ is the Fock vacuum annihilated by all the $\ch_j$, and $\mathcal{N}$ is a normalization
factor.

To compute the normalization factor $\mathcal{N}$ we use the fact that, for any complex antisymmetric 
matrix $g$, there exists a unitary matrix $W$ such that
\beq
	W^T g W = \Lambda\ ,
\eeq 
where 
\beq
	\Lambda = \text{diag}\left\{\begin{pmatrix} 
	0 & \lambda_1 \\
	-\lambda_1 & 0 
	\end{pmatrix},
	\begin{pmatrix} 
	0 & \lambda_2 \\
	-\lambda_2 & 0 
	\end{pmatrix},\dots,
	\begin{pmatrix} 
	0 & \lambda_{L/2} \\
	-\lambda_{L/2} & 0 
	\end{pmatrix}
	\right\}\ ,
\eeq
and where $\lambda_{\al}\in\mathbb{R}$, $\lambda_{\al} \geq 0$ for $\al=1,\dots,L/2$.
Note here that we assume that $L$ is even. Let us also label the columns of $W$ by 
$w_1,\td{w}_1,w_2,\td{w}_2,\dots,w_{L/2},\td{w}_{L/2}$. An additional property of the matrix
$W$, which follows from its explicit construction, is that
$w_{\al}$ is orthogonal to $\td{w}_{\al}$ for any $\al$, i.e., 
$(w_{\al},\td{w}_{\al})=0$ where $(x,y)=\sum_{i=1}^L \ov{x}_i y_i$ denotes the standard inner product on 
$\mathbb{C}^L$.

This theorem allows us to write
\beq
	\hat{G}=\sum_{\al=1}^{L/2}\lambda_{\al}\hat{C}^{\dg}_{\al}\hat{\td{C}}^{\dg}_{\al}\ ,
\eeq
where we defined 
\begin{subequations}
\beqa
	\hat{C}_{\al}&=& \sum_{i=1}^L w_{\al,i} \hat{c}_i	\\
	\hat{\td{C}}_{\al}&=& \sum_{i=1}^L \td{w}_{\al,i} \hat{c}_i	\ ,
\eeqa
\end{subequations}
and $w_{\al,i}$ is the $i^{th}$ component of the vector $w_{\al}$.
The fact that $(w_{\al},\td{w}_{\al})=0$ for all $\al$ then implies that these operators are a
set of $L$ independent complex fermions. In particular, a fermion operator without a tilde anticommutes
with any fermion operator with a tilde, for example $\{\hat{C}_{\al},\hat{\td{C}}^{\dg}_{\beta}\}=0$.
The remaining anticommutation relations are standard, for example  
$\{\hat{C}_{\al},\hat{C}_{\beta}\}=0$ and $\{\hat{C}_{\al},\hat{C}^{\dg}_{\beta}\}=\delta_{\al\beta}$.

This result allows us to rewrite the state $|+\ran$ in the simple ``BCS-like'' form
\beqa
	|+\ran &=& \mathcal{N}e^{\sum_{\al=1}^{L/2}\lambda_{\al}\hat{C}^{\dg}_{\al}\hat{\td{C}}^{\dg}_{\al}}|0\ran \nnb \\
	&=& \mathcal{N}\prod_{\al=1}^{L/2}\left[1+\lambda_{\al}\hat{C}^{\dg}_{\al}\hat{\td{C}}^{\dg}_{\al}\right]|0\ran\ .
\eeqa
An elementary computation now shows that 
\beq
	\lan +|+\ran= |\mathcal{N}|^2 \prod_{\al=1}^{L/2}\left[1+\lambda_{\al}^2\right]\ ,
\eeq
and so we can choose
\beq
	|\mathcal{N}|^{-2}= \prod_{\al=1}^{L/2}\left[1+\lambda_{\al}^2\right]\ . \label{eq:norm}
\eeq

Next, we compute the generating function $\lan +|e^{\tau \hat{N}_w}|+\ran$ 
for expectation values of the particle number operator
$\hat{N}_w=\sum_j \ch^{\dg}_j\ch_j$ in the state $|+\ran$. We have
\beqa
	\lan +|e^{\tau \hat{N}_w}|+\ran &=& \lan +|e^{\frac{\tau}{2} \hat{N}_w}e^{\frac{\tau}{2} \hat{N}_w}|+\ran \nnb \\
	&=& |\mathcal{N}|^2 \lan 0|e^{\frac{\tau}{2} \hat{N}_w}e^{-\frac{\tau}{2} \hat{N}_w}e^{\hat{G}^{\dg}}e^{\frac{\tau}{2} \hat{N}_w}e^{\frac{\tau}{2} \hat{N}_w}e^{\hat{G}}e^{-\frac{\tau}{2} \hat{N}_w}e^{\frac{\tau}{2} \hat{N}_w}|0\ran \nnb \\
	&=& |\mathcal{N}|^2 \lan 0|e^{\frac{1}{2}\sum_{i,j=1}^L e^{\tau}\ov{g}_{ij}\ch_j\ch_i}e^{\frac{1}{2}\sum_{i,j=1}^L e^{\tau}g_{ij}\ch^{\dg}_i\ch^{\dg}_j}|0\ran \nnb \\
	&=& |\mathcal{N}|^2 \prod_{\al=1}^{L/2}\left[1+e^{2\tau}\lambda_{\al}^2\right] \nnb \\
	&=& \prod_{\al=1}^{L/2}\left[\frac{1+e^{2\tau}\lambda_{\al}^2}{1+\lambda_{\al}^2}\right]\ .
\eeqa
We can use this generating function to compute the average number of particles 
$\ov{N}_{w,+} := \lan + |\hat{N}_w|+\ran$ in the
state $|+\ran$, and we find that
\beq
	\ov{N}_{w,+} = 2\sum_{\al=1}^{L/2}\frac{\lambda_{\al}^2}{1+\lambda_{\al}^2}\ . 
\eeq
In particular, it is clear that $0\leq \ov{N}_{w,+} \leq L$.

We now derive an upper bound on the expectation value
$\lan +| (\hat{N}_w-\ov{N}_{w,+})^2|+\ran$, as this upper bound is a crucial ingredient
in our proofs of Theorems 3, 4 and 5. To start, we define the new
generating function
\beq
	F_{1,+}(\tau)= \lan +|e^{\tau (\hat{N}_w-\ov{N}_{w,+})}|+\ran\ ,
\eeq
and in terms of this function we have
\beq
	\lan +| (\hat{N}_w-\ov{N}_{w,+})^p|+\ran= \frac{d^p F_{1,+}(\tau)}{d\tau^p}\Bigg|_{\tau=0}\ .
\eeq
This function satisfies $F_{1,+}(0)=1$ and $F'_{1,+}(0)=0$, where the prime denotes a derivative
with respect to $\tau$. These two properties imply that the first few derivatives of $F_{1,+}(\tau)$, when
evaluated at $\tau=0$, can be expressed simply in terms of derivatives (also evaluated at 
$\tau=0$) of a new function
\beq
	F_{2,+}(\tau)= \ln\left[F_{1,+}(\tau)\right]\ .
\eeq
For example, we find that
\begin{subequations}
\beqa
	F''_{1,+}(0) &=& F''_{2,+}(0) \\
	F'''_{1,+}(0) &=& F'''_{2,+}(0) \\
	F''''_{1,+}(0) &=& F''''_{2,+}(0) + 3[F''_{2,+}(0)]^2\ .
\eeqa
\end{subequations}
We only need the second derivative of $F_{2,+}(\tau)$ to compute $\lan +| (\hat{N}_w-\ov{N}_{w,+})^2|+\ran$, 
and we find that
\beq
	F''_{2,+}(0)= 4\sum_{\al=1}^{L/2}\frac{\lambda_{\al}^2}{(1+\lambda_{\al}^2)^2}\ .
\eeq
We can obtain an upper bound on this quantity by finding the maxima of the function appearing in the sum 
for $\lambda_{\al}^2\in[0,\infty)$. The function $\frac{4\lambda_{\al}^2}{(1+\lambda_{\al}^2)^2}$
takes on its maximum value of $1$ at $\lambda_{\al}^2= 1$, and so 
$F''_{2,+}(0)\leq \sum_{\al=1}^{L/2} 1 =\frac{L}{2}$. Thus, we find the bound
\beq
	\lan +| (\hat{N}_w-\ov{N}_{w,+})^2|+\ran \leq \frac{L}{2} \label{eq:thm1-bound-even}\ .
\eeq

We now turn to the derivation of the analogous bound for the odd parity ground state, which we denote
by $|-\ran$. This state is constructed by acting on $|+\ran$ with the creation operator $\hat{d}^{\dg}_i$ 
with the lowest energy $E_i$. We adopt the convention that this lowest energy is achieved for $i=1$, so that
\beq
	|-\ran= \hat{d}^{\dg}_1|+\ran\ .
\eeq  
Note that this state is normalized, $\lan -|-\ran=1$, since $\hat{d}_1|+\ran=0$. 
By using the identity 
\beq
	e^{-\hat{G}}\hat{d}^{\dg}_1 e^{\hat{G}}= \hat{d}^{\dg}_1+\sum_{j=1}^L (\ov{v}g)_{1j}\hat{c}^{\dg}_j\ ,
\eeq
where $\ov{v}$ is the complex conjugate of the matrix $v$, we find that $|-\ran$ can be rewritten in the form 
\beq
	|-\ran= \mathcal{N}e^{\hat{G}}\sum_{j=1}^L\left[\ov{u}_{1j}+(\ov{v}g)_{1j}\right]\hat{c}^{\dg}_j|0\ran\ .
\eeq
The key property that we now use is that, since the matrix $W$ is unitary, it is possible to rewrite 
any operator $\ch_j$ as a linear combination of the new complex fermions 
$\hat{C}_{\al}$ and $\hat{\td{C}}_{\al}$. This implies the existence of coefficients $X_{\al}$ and
$\td{X}_{\al}$ such that 
\beq
	\sum_{j=1}^L\left[\ov{u}_{1j}+(\ov{v}g)_{1j}\right]\hat{c}^{\dg}_j= \sum_{\al=1}^{L/2}\left( X_{\al}\hat{C}^{\dg}_{\al} +\td{X}_{\al}\hat{\td{C}}^{\dg}_{\al}\right)\ .
\eeq
It turns out that we do not actually need to know the exact expressions for $X_{\al}$ and
$\td{X}_{\al}$ for the calculations in the rest of this section. In terms of these
coefficients, we find that $|-\ran$ can be rewritten as
\beqa
	|-\ran &=& \mathcal{N}e^{\hat{G}}\sum_{\al=1}^{L/2}\left( X_{\al}\hat{C}^{\dg}_{\al} +\td{X}_{\al}\hat{\td{C}}^{\dg}_{\al}\right)|0\ran \nnb \\
	&=& \mathcal{N}\sum_{\al=1}^{L/2}\left( X_{\al}\hat{C}^{\dg}_{\al} +\td{X}_{\al}\hat{\td{C}}^{\dg}_{\al}\right)\prod_{\beta\neq\al}\left[1+\lambda_{\beta}\hat{C}^{\dg}_{\beta}\hat{\td{C}}^{\dg}_{\beta}\right]|0\ran\ ,
\eeqa
where we expanded $e^{\hat{G}}= \prod_{\beta=1}^{L/2}\left[1+\lambda_{\beta}\hat{C}^{\dg}_{\beta}\hat{\td{C}}^{\dg}_{\beta}\right]$ and used the fact that all complex fermion operators square to zero.

We now use our new expression for $|-\ran$ to compute the norm $\lan -|-\ran$, and we find that
\beqa
	\lan-|-\ran &=& |\mathcal{N}|^2 \sum_{\al=1}^{L/2}\left( |X_{\al}|^2 +|\td{X}_{\al}|^2\right)\prod_{\beta\neq\al}\left[1+\lambda_{\beta}^2\right] \nnb \\
	&=& |\mathcal{N}|^2 \prod_{\beta=1}^{L/2}\left[1+\lambda_{\beta}^2\right] \sum_{\al=1}^{L/2}\frac{\left( |X_{\al}|^2 +|\td{X}_{\al}|^2\right)}{1+\lambda_{\al}^2} \nnb \\
	&=& \sum_{\al=1}^{L/2}\frac{\left( |X_{\al}|^2 +|\td{X}_{\al}|^2\right)}{1+\lambda_{\al}^2}\ ,
\eeqa
where we used Eq.~\eqref{eq:norm} for $\mathcal{N}$. Since we already know that $\lan -|-\ran=1$, this
expression yields the important formula
\beq
	\sum_{\al=1}^{L/2}\frac{\left( |X_{\al}|^2 +|\td{X}_{\al}|^2\right)}{1+\lambda_{\al}^2}=1\ . \label{eq:important-norm-formula}
\eeq
This formula will be crucial for the bound that we derive in the rest of this section. It is also the only 
piece of information about the coefficients $X_{\al}$ and $\td{X}_{\al}$ that we actually need for our 
calculations.

Using our new form for $|-\ran$, we can also calculate the generating function 
$\lan-|e^{\tau \hat{N}_w}|-\ran$. Using manipulations similar to those used in the derivation of the 
expression for $\lan -|-\ran$, we find that
\beq
	\lan-|e^{\tau \hat{N}_w}|-\ran	= \lan+|e^{\tau \hat{N}_w}|+\ran e^{\tau} Q(\tau) \ ,
\eeq
where we defined the function
\beq
	Q(\tau)=	\sum_{\al=1}^{L/2}\frac{\left( |X_{\al}|^2 +|\td{X}_{\al}|^2\right)}{1+e^{2\tau}\lambda_{\al}^2}\ .
\eeq
Note also that Eq.~\eqref{eq:important-norm-formula} implies that $Q(0)=1$ (one can also see that
$Q(\tau)\leq 1$ for $\tau\geq 0$). 
If we define the average $\ov{N}_{w,-} := \lan - |\hat{N}_w|-\ran$ like in
the even parity case, then we can define the generating function for expectation values of 
powers of $\hat{N}_w-\ov{N}_{w,-}$,
\beq
	F_{1,-}(\tau)= \lan -|e^{\tau (\hat{N}_w-\ov{N}_{w,-})}|-\ran\ ,
\eeq
and we find that this can be written as
\beq
	F_{1,-}(\tau)= e^{(1+\ov{N}_{w,+}-\ov{N}_{w,-})\tau}F_{1,+}(\tau)Q(\tau)\ .
\eeq
By construction, this generating function has the same properties $F_{1,-}(0)=1$ and $F_{1,-}'(0)=0$ as
$F_{1,+}(\tau)$. We are therefore led to again define a second generating function $F_{2,-}(\tau)$ 
equal to the logarithm of $F_{1,-}(\tau)$, and we find that
\beqa
	F_{2,-}(\tau) &=& \ln\left[F_{1,-}(\tau)\right] \nnb \\
	&=& (1+\ov{N}_{w,+}-\ov{N}_{w,-})\tau + F_{2,+}(\tau) + \ln\left[Q(\tau)\right]\ .
\eeqa
We can now compute the desired expectation values of powers of $\hat{N}_w-\ov{N}_{w,-}$ by taking 
derivatives of this second generating function, just as in the even parity case.

The expectation value that we need for the proofs of Theorems 3, 4 and 5 is
\beq
	\lan -|(\hat{N}_w-\ov{N}_{w,-})^2|-\ran= F_{2,-}''(0)\ .
\eeq
Using our explicit expression for $F_{2,-}(\tau)$ from above, and using $Q(0)=1$, we find that 
\beqa
	F_{2,-}''(0) &=& F_{2,+}''(0)+ Q''(0)- \left[Q'(0)\right]^2 \nnb \\
	 &\leq&  F_{2,+}''(0)+ Q''(0)\ ,
\eeqa
where the second line follows immediately from the fact that $\left[Q'(0)\right]^2$ is positive. We
already know that $F_{2,+}''(0)\leq L/2$, and so the only work that remains
is to find an upper bound for $Q''(0)$. The
explicit expression for $Q''(0)$ is
\beqa
	Q''(0) &=& \sum_{\al=1}^{L/2}\left( |X_{\al}|^2 +|\td{X}_{\al}|^2\right)\left(\frac{8\lambda_{\al}^4}{(1+\lambda_{\al}^2)^3}-\frac{4\lambda_{\al}^2}{(1+\lambda_{\al}^2)^2}\right) \nnb \\
	 &\leq& \sum_{\al=1}^{L/2}\left( |X_{\al}|^2 +|\td{X}_{\al}|^2\right)\left(\frac{8\lambda_{\al}^4}{(1+\lambda_{\al}^2)^3}\right)\ ,
\eeqa
where the second line follows after dropping the negative term from the first line. Next, we use the
simple inequality
\beq
	\frac{\lambda_{\al}^2}{1+\lambda_{\al}^2} \leq 1 \label{eq:simple}
\eeq
in the remaining term to find that
\beq
	Q''(0) \leq \sum_{\al=1}^{L/2}\left( |X_{\al}|^2 +|\td{X}_{\al}|^2\right)\left(\frac{8}{1+\lambda_{\al}^2}\right) = 8Q(0) = 8\ .
\eeq
Thus, we find that 
\beq
	\lan -|(\hat{N}_w-\ov{N}_{w,-})^2|-\ran \leq \frac{L}{2}+8 \label{eq:thm1-bound-odd}\ ,
\eeq
and this is the last bound that we need for our proofs. 
Finally, it is important to note here that the leading term in this bound 
(the term with the largest power of $L$) exactly matches the leading term in the analogous bound for the even 
parity ground state.


%